\documentclass{aa}
\usepackage{graphicx}
\newcommand{\lsim}{\raisebox{-5pt}{$\;\stackrel{\textstyle <}{\sim}\;$}}
\newcommand{\gsim}{\raisebox{-5pt}{$\;\stackrel{\textstyle >}{\sim}\;$}}

\begin{document}
   \title{  The dust obscuration bias in  Damped Lyman\,$\alpha$ systems
}
 
   \author{Giovanni Vladilo 
          \inst{1}
          \and
          C\'eline P\'eroux 
          \inst{2} 
          }

   \offprints{G. Vladilo}

   \institute{
              Osservatorio Astronomico di Trieste - Istituto Nazionale di Astrofisica, 
              Via Tiepolo 11, 34131 Trieste, Italy. \\
              \email{vladilo@ts.astro.it}
  \and
             European Southern Observatory, Karl-Schwarzschild-Str. 2,
	85748 Garching-bei-M\"unchen, Germany. \\
             \email{cperoux@eso.org}
                }

   \date{Received ...; accepted ...}
   
 \titlerunning{Dust obscuration bias in  Damped Ly\,$\alpha$ systems}
 \authorrunning{G. Vladilo and C. P\'eroux }

 \abstract{  
 We present a new study of the effects 
of quasar obscuration  on the
statistics of Damped Ly\,$\alpha$ (DLA)  systems. 
We show that the extinction
of   any Galactic or extragalactic \ion{H}{i} region,
 $A_\lambda$,    increases
  linearly with the column density of zinc, 
with a 
turning point 
$\partial A_\lambda / \partial (\log N_\mathrm{Zn}) =1$, 
above which background sources are suddenly obscured.  
%
We derive a relation 
$A_\lambda=A_\lambda(N_\mathrm{H},Z,z)$
between the extinction of a DLA system and its     
\ion{H}{i} column density, $N_\mathrm{H}$, metallicity, $Z$, 
fraction of iron in dust, $f_\mathrm{Fe}(Z)$, and redshift, $z$. 
From this relation we estimate the fraction of  DLA  systems
    missed as a consequence of their own   extinction
in magnitude-limited surveys. 
%
We derive a method for recovering the true
frequency distributions of $N_\mathrm{H}$ and $Z$ in DLAs, 
$f_{N_\mathrm{H}}$ and $f_Z$,  
using the biased distributions measured in the redshift range
where the observations have sufficient statistics ($1.8 \leq z \leq 3$). 
%
By applying our method   we find that the
   well-known empirical thresholds  of   DLA  column densities,
$N_\mathrm{Zn} \la 10^{13.1}$ atoms cm$^{-2}$ and
$N_\ion{H}{i} \la 10^{22}$ atoms cm$^{-2}$, can be 
successfully explained in terms of the obscuration effect
without tuning of the local dust parameters. 
%
The obscuration has a modest effect on the distribution of quasar apparent magnitudes,
but plays an important role in shaping the statistical distributions of DLAs.
%
The exact estimate of the bias  is still
limited by the  paucity of the data  
 ($\simeq$ 40  zinc measurements at $1.8 \leq z \leq 3$). 
We find that the fraction of DLAs missed  
as a consequence of   obscuration is $\sim$ 30\% to 50\%, 
   consistent  with the results of  
 surveys of  radio-selected quasars.
By modelling the   
metallicity distribution with a Schechter function
we find that the mean metallicity  can be
$\sim 5$ to 6 times higher than the  value  
commonly reported for DLAs at $z \sim 2.3$.
 \keywords{ ISM: dust, extinction -- Galaxies: ISM -- Galaxies: high-redshift
 -- Quasars: absorption lines  }
  }

\maketitle

   
%
 
\section{Introduction}

Quasar absorption line systems probe the diffuse gas in the Universe
over cosmological scales and at various stages of evolution. 
Among these absorbers, the Damped Ly$\alpha$ systems (hereafter DLAs) 
have the highest \ion{H}{i} column
densities ($N_\ion{H}{i} \geq 2 \times 10^{20}$ atoms cm$^{-2}$)
and  are believed to originate in \ion{H}{i} regions of high-redshift galaxies  
(Wolfe et al. 1986). 
Also, they have   higher metallicities\footnote{
We adopt
the usual definition [X/H] = log (X/H) - log (X/H)$_{\sun}$.}  than
any other class of quasar absorbers   ([M/H] $\approx -1.1$
dex; Pettini et al. 1994) and as such are expected to contain  dust.
%
If present, the  dust in DLAs  will absorb and scatter the radiation
of the background quasar, dimming   its apparent magnitude
("extinction" effect) and changing the slope of its spectral distribution
("reddening"). 
In magnitude limited surveys, the extinction may obscure the quasar,
leading to an observational selection effect known  as
obscuration bias (Ostriker \& Heisler, 1984; Fall \& Pei 1989, 1993). 


The bias can lead to misleading conclusions on the nature
of the galaxies associated with the DLA systems (herefater DLA galaxies),
because the \ion{H}{i} regions most enriched by metals and dust would
be systematically missed in the surveys.  
In addition,  some properties
of the high-redshift Universe that can only be derived from studies of DLAs,
would also be affected by the bias. 
One example is
the  comoving mass density of neutral gas,
$\Omega_\mathrm{DLA}$, and its evolution with redshift,
which is an indicator of gas consumption due to
 star formation (Wolfe et al. 1995). 
Another example is the mean cosmic metallicity
inferred from abundance studies of DLAs (Pettini et al. 1997),
which is related to  the star formation rate 
in the early Universe
(Pei \& Fall 1995, hereafter PF95; Pei et al. 1999).
 
Whether the obscuration effect is important, depends 
on the actual amount of dust in DLAs. 
The first evidence of dust, in the form of reddening,
was reported by Pei et al. (1991). 
The complex variability 
of the  quasar continuum  makes it impossible to determine the
reddening in individual cases.  However, 
from a statistical
comparison of quasars with and without foreground DLAs
one can search for a systematic change of the continuum slope.
In this way,   Pei et al. found a systematic difference,  
suggestive of reddening,
using a simplified power-law fit to the continuum distribution.

%

The detection or reddening motivated detailed studies    
of quasar obscuration. 
Fall \& Pei (1993;  hereafter FP93)  presented an analytical method of computing
the obscuration   from the observed luminosity function
of quasars, the typical dust-to-gas ratio of DLAs and the empirical
distribution of \ion{H}{i} column densities. 
By modelling the radial distribution of the neutral gas in DLA galaxies, 
they concluded  that a large fraction of quasars may be obscured
in optically selected samples.   
%
%
The large uncertainty   of the reddening measurements prevented
  reaching firm conclusions on the dust-to-gas ratios and
on the magnitude of the bias. 

An alternative approach to measure the amount of the  dust  in DLAs   
consists in measuring the differential depletions between
refractory and volatile elements 
(Pettini et al. 1994;  Hou et al. 2001; Prochaska \& Wolfe 2002;
Ledoux et al. 2003; Vladilo 2002, 2004).
%
%
%
The  depletions can be converted into dust-to-gas ratios, yielding
an approximate estimate of  the quasar extinction  
(Vladilo et al. 2001b; Prochaska \& Wolfe 2002).   
This method, as well as the reddening measurements,
only   give information on the detected DLAs, leaving
open the possibility that more dusty \ion{H}{i} regions
remain undetected. 

%

 
A more direct probe of the obscuration effect is a comparison
of DLA statistics in optically selected and radio selected quasar
samples, given that the latter are unaffected by dust extinction.
Ellison et al. (2001) compiled a homogeneous sample of radio selected
quasars and searched for DLAs towards every target, irrespective of
its optical magnitude (CORALS survey). They concluded that dust-induced bias in
previous magnitude limited surveys may have led to underestimating the
\ion{H}{i} mass density, $\Omega_\mathrm{DLA}$, and the number
density per unit redshift interval, $n(z)$, of DLAs by at most a
factor of two at $z \sim 2.3$. 
In addition, they found tentative evidence that $n(z)$ is greater in
fainter quasars, as expected by the obscuration effect.
In a follow-up  study of CORALS metallicities,
Akerman et al. (2005) find 
$ [ < \mathrm{Zn/H} > ] = -0.88 \pm 0.21 $. This value is  
   higher  than in a 
control sample, 
as expected by the obscuration, but only at $1 \sigma$ level. 
In an extension of the CORALS survey at $0.6 < z < 1.7$
 Ellison et al. (2004)
find  $n(z) = 0.16 ^{+0.08}_{-0.06}$
using MgII absorbers as DLA candidates.
 This value is  consistent with magnitude-limited estimates
at the same $z$, but the $1 \sigma$ error permits a factor of 2.5
difference in the sense predicted by the obscuration. 

The results of the CORALS survey are not yet conclusive since the
original sample of Ellison et al. had limited statistics, in particular at the
high values of $N(\ion{H}{i})$ where the effect is expected to be more
important. This weak evidence of obscuration, 
together with the non-detection of reddening in a large sample
of quasars with and without candidate DLAs, 
recently reported by Murphy \& Liske (2004), 
are conveying the impression that the obscuration effect
is not  important.

Yet, indirect evidence for the existence of the bias  comes  
from other studies.
 The analysis of
[Zn/H] versus $\log N(\ion{H}{i})$ by Boiss\'e et al. (1998) showed
that none of the 
DLAs with zinc detection known at the time have
a metallicity above the threshold [Zn/H] $+\log N(\ion{H}{i}) \simeq
20.5$, corresponding to $N(\ion{Zn}{ii}) \simeq 10^{13.1}$ atoms
cm$^{-2}$. They proposed that this threshold could be attributed to
obscuration effect, assuming that metal column density tracks the dust
column density. 

While this assumption has not been proven by subsequent investigations,
the  threshold proposed by Boiss\'e et al.  
has been invoked
to reconcile  several predictions of galactic evolution models 
(Prantzos \& Boissier 2000; Hou et al. 2001)
and cosmological simulations 
(Cen et al. 2003) with the observed properties of DLAs. 

Another potential evidence of the bias is the lack of DLAs with \ion{H}{i} column densities $N(\ion{H}{i}) \ga 10^{22}$ atoms cm$^{-2}$. 
In this case, the obscuration   has been invoked to
explain  the significant fraction of model galaxy disks 
predicted to have higher \ion{H}{i} column densities  (Churches et al. 2004). 
As an alternative explanation,
it has been proposed that  hydrogen may undergo a sudden transition from 
the atomic
to the molecular form  above a critical
column density threshold  (Schaye 2001), in which
case the absorber would not be detected as a DLA system. 
 

In the present work  we address several open questions
concerning the existence and   importance of the obscuration bias. 
In Section 2, 
we   perform a careful investigation of all the factors
that determine the extinction of a Galactic or extragalactic \ion{H}{i}
region. 
%
In     Section 3 we   report some indirect evidence 
of obscuration effects.
%
In Section 4 we present a  
  mathematical formulation
 aimed at  recovering  the frequency distributions
of  \ion{H}{i} column densities and metallicities of DLAs
 from the    
observed  distributions affected by obscuration bias. 
The first implementation of this method is described
in Section 5 and the results are   discussed in Section 6. 

\section{ The relation between $N_\mathrm{Zn}$ and extinction   }

%
We adopt  
the simplified notation $N_\mathrm{Zn} = N(\ion{Zn}{ii})$ and
$N_\mathrm{H} = N(\ion{H}{i})$, 
ignoring contributions from ionization states other than the
dominant one (see Vladilo et al. 2001a and references therein). 
 
We start by considering  a refractory and a volatile\footnote{
Refractory elements easily condense
into dust form, while volatile elements tend to remain in the gas phase.
}    
element, $\mathrm{E_r}$ and $\mathrm{E_v}$,  
with   interstellar abundance ratio by number 
  $(  \mathrm{E_r} /  \mathrm{E_v}   )$.  
The ratio of the total column densities (gas plus dust) 
of these two elements along an interstellar line of sight will be 
$
 N^\mathrm{tot}_\mathrm{r} / N^\mathrm{tot}_\mathrm{v}  = 
 ( \mathrm{E_r} /  \mathrm{E_v}  ).
$
We call $N^\mathrm{d}_\mathrm{E}$ 
and $N_\mathrm{E}$ the column densities of an element E in the dust  
and  in the gas, respectively.  The fraction of atoms of E in dust form will be
$f_{\mathrm{E}} 
\equiv N^\mathrm{d}_{\mathrm{E}}/N^\mathrm{tot}_{\mathrm{E}}$. 
From these definitions we have 
$ N^\mathrm{tot}_{\mathrm{r}} 
= N^\mathrm{d}_\mathrm{r} / f_{\mathrm{r}}$ and 
$N^\mathrm{tot}_\mathrm{v} 
= N_\mathrm{v} / (1-f_\mathrm{v})$.
Combining these relations we have 
\begin{equation}
N^\mathrm{d}_{\mathrm{r}} 
=  \, { f_\mathrm{r} \over (1-f_\mathrm{v})} \,
 \left( { {\mathrm{E_r} \over \mathrm{E_v} } } \right)  \,  N_\mathrm{v} 
 ~.
\label{RoverV}
\end{equation} 

We next consider the relation between  $N^\mathrm{d}_\mathrm{r}$ and
the column density of dust grains,
$N_\mathrm{gr}$, i.e. the 
number of grains per cm$^{2}$ along the line of sight. 
For spherical grains of radius $r_\mathrm{gr}$ (cm)
and internal mass density $\varrho_\mathrm{gr}$ (g cm$^{-3}$)
it is easy to show that 
\begin{equation}
N^\mathrm{d}_{\mathrm{r}} \, \mathrm{A_r} \, m_\mathrm{H} =
{4 \over 3 } \, \pi 
 ~ r_\mathrm{gr}^{3} ~ \varrho_\mathrm{gr}  ~   N_\mathrm{gr} ~ X^\mathrm{d}_\mathrm{r}
  ~,
\label{Rgrains}
\end{equation}
where    $\mathrm{A_r}$ is the atomic mass of the refractory element and
$X^\mathrm{d}_\mathrm{r}$ the mass fraction
in the dust of the same element. 
 
Finally, we consider the relation between  the interstellar extinction in magnitudes
 at the wavelength $\lambda$, $A_\lambda$,   
 and the column density of dust grains. 
For spherical grains  with extinction efficiency factor $Q_{e,\lambda}$ 
(see Spitzer 1978, Chapter 7) this relation is  
\begin{equation}
A_\lambda  = 1.086 ~ 
  N_\mathrm{gr} \,\,  Q_{e,\lambda} \,\, \pi \, r_\mathrm{gr}^2 ~.
\label{AVgrains}
\end{equation} 
 
By combining Eqs. (\ref{RoverV}), (\ref{Rgrains}) and (\ref{AVgrains})
we derive 
\begin{equation}
A_\lambda  = a_\lambda ~  N_{\mathrm{v}} ~,
\label{BigA_Nv}
\end{equation}
where we define
\begin{equation}
a_\lambda =  1.817 \times 10^{-24} ~ \mathrm{A_r} ~ g_\lambda  ~
 { f_\mathrm{r} \over (1-f_\mathrm{v})}  \,
 \left( { {\mathrm{E_r} \over \mathrm{E_v} } } \right) 
   ~,
\label{SmallA_Nv}
\end{equation}
and we isolate in the term
\begin{equation}
g_\lambda = s ~
{ Q_{e,\lambda}  \over 
r_\mathrm{gr}  \, \varrho_\mathrm{gr} \,   
X^\mathrm{d}_\mathrm{r}  } ~
\label{glambda}
\end{equation}
all the   properties of the grains (shape, extinction efficiency,
size, density, and composition). For spherical grains $s = 3/4$.

The factor 
${ f_\mathrm{r} / (1-f_\mathrm{v})}$
is approximately constant  by definition of volatile and refractory elements. 
 In a medium of constant composition, such as the local ISM, also the abundance $\left( { {\mathrm{E_r} / \mathrm{E_v} } } \right)$
is constant. 
Therefore, for a given type of dust,   the term $a_\lambda$ is constant  
and the expression (\ref{BigA_Nv}) implies   that   {\em the interstellar
extinction   increases linearly with the gas-phase column density of any volatile element
measured in the same line of sight}. 
%

Differences in the shape, size, density, composition and extinction efficiency
of the grains will induce a scatter in the relation.  
The dependence of $g_\lambda$ on the grain size  $r_\mathrm{gr}$ is linear,  
while in Eqs. (\ref{Rgrains}) and (\ref{AVgrains}) it was cubic and quadratic.
%
Since the grain size distribution
is   a key factor in determining the wavelength dependence  of the 
interstellar extinction (see e.g. Draine 2003), the mild dependence of
$g_\lambda$ on $r_\mathrm{gr}$ suggests   that
{\em  the
relation between $A_\lambda$ and $N_\mathrm{v}$ may be similar
even in interstellar regions with different types of
extinction curves}.

\begin{figure}
   \centering
 \includegraphics[width=7.5cm,angle=0]{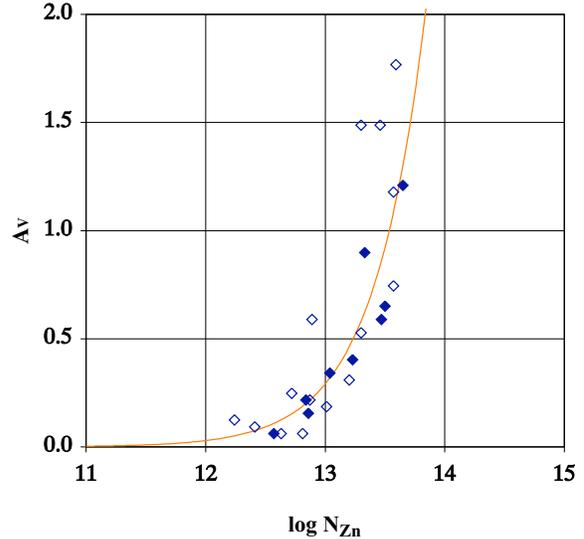}
\caption{Extinction $A_V=R_V \,  E_{B-V}$ (mag)
versus \ion{Zn}{ii} column density (atoms cm$^{-2}$)
in Galactic interstellar lines of sight. Filled diamonds: 
HST data 
from  Roth \& Blades (1995). Empty diamonds: 
IUE data from Van Steenberg \& Shull (1988). 
Curve: extinction law (\ref{AV_NZn}) with    
$ a_V = 0.29 \times 10^{-13}$ mag cm$^2$ derived empirically as
explained in the text. }
\label{figAV_NZn}
\end{figure}

By adopting zinc as the volatile element       and considering the   extinction in    
 the $V$ band ($\lambda_V=0.55 \mu$m), we obtain from (\ref{BigA_Nv})
  the relation
\begin{equation}
A_V  \simeq a_V 
\,   N_\mathrm{Zn} ~,
\label{AV_NZn}
\end{equation}
where the parameter $a_V$ (mag cm$^2$)
can be determined  by comparing measurements of $A_V$ and  $N_\mathrm{Zn}$  
in individual lines of sight.   
In this way it is possible to  measure the scatter of $a_V$  
due to variations of the grain properties. 

To estimate $a_V$ we gathered 
the $N_\mathrm{Zn}$ measurements 
obtained from    HST    (Roth \& Blades 1995) and  
    IUE (Van Steenberg \& Shull 1988) high-resolution spectra.  
The IUE data are more numerous, but
have lower  resolution and signal-to-noise ratio.  
To reduce the impact of these differences we selected the IUE
measurements  with total error $< 0.3$ dex 
  in $\log N_\mathrm{Zn}$.
To minimize the contamination 
by stellar absorptions on top of the IUE \ion{Zn}{ii} interstellar lines
(Van Steenberg \& Shull 1988), we selected
only IUE spectra of fast-rotating stars ($V \sin i \geq$ 200 km s$^{-1}$). 
We then scaled the IUE  column densities to match the 
Bergeson \& Lawler (1993)
\ion{Zn}{ii} oscillator strengths adopted in the HST data 
and in the current research.  
As far as the extinctions are concerned, we used the   $A_V$  
values given by Guarinos (1991). 
As a result, we obtained a sample of 11 lines of sight.
The mean value of extinction per unit zinc column density that we derive
is $<\!\! a_V \!\!> \, = 0.27 (\pm 0.06) \, 10^{-13}$ mag cm$^2$ and
$0.38 (\pm 0.20) \, 10^{-13}$ mag cm$^2$ for the HST and IUE sub-samples
(4 and 7 lines of sight, respectively).  

In order to enlarge the sample we also considered the color excess
$E_{B-V}$,  which is more commonly  measured than $A_V$.
The ratio of general-to-selective extinction,
$
R_V = { A_V / E_{B-V}}
$,
is  approximately constant,  with a mean value
$<\!\! R_V \!\!> \, \simeq 3.1$ in the diffuse ISM of the Milky Way
(see e.g. Draine 2003).
We used this property to derive a mean value 
$<\!\! a_V \!\!> \, = \,<\!\! R_V \!\!> \, <\!\! E_{B-V}/N_\mathrm{Zn} \!\!>$ 
from reddening and zinc column density measurements. 
The $E_{B-V}$  data 
were taken from the
extinction catalog of Savage et al. (1985).   
The resulting sample includes 24 lines of sight, shown in Fig. \ref{figAV_NZn}. 
The mean value 
of the HST sub-sample (9 lines of sight) is
$<\!\! a_V \!\!> \, = 3.1 \times 0.089 \, (\pm 0.030) \, 10^{-13}$ mag cm$^2$,
and of the IUE sub-sample (15 lines of sight),
$3.1 \times 0.123 \, (\pm 0.071) \, 10^{-13}$ mag cm$^2$.
%
We did not find significant correlations  between
$<\!\! E_{B-V}/N_\mathrm{Zn} \!\!>$ and the parameters of
UV extinction provided by   Savage et al. (1985),
such as  the strength of the emission bump
or the steepness of the UV rise. 
The lack of these correlations gives us a justification for averaging
the results obtained from   lines of sight with
  different types of dust. 
 
The results obtained from the $E_{B-V}$ analysis  
are identical to those obtained from the more limited
sample of direct measurements of $A_V$. 
The larger dispersion of the IUE
sub-sample is probably due to the lower quality of the data.
%
%
We adopted the weighted average of the
IUE and HST sub-samples,   which yields 
$<\!\! a_V \!\!> \, = 0.29 (\pm 0.07) \, 10^{-13}$ mag cm$^2$,
a value almost identical to that of the HST sub-sample. 
%
The   relation   (\ref{AV_NZn}) obtained from this estimate of  $a_V$
is displayed in Fig. \ref{figAV_NZn} (solid curve). The exponential rise
of the extinction is due to the fact that the \ion{Zn}{ii} column density
is plotted on a logarithmic scale. 
In this scale the rise of the extinction is mild 
at low values of  $\log N_\mathrm{Zn}$, but very fast at high values. 
The turning point 
marking the transition between these two regimes
can be estimated from the condition        
$\partial A_V / \partial (\log N_\mathrm{Zn}) =1$.
This criterion yields  $\log N_\mathrm{Zn} 
= - \log (<\!\! a_V \!\!> \, \ln 10) = 13.18$. 
{\em Above this turning point 
 the extinction undergoes a dramatic rise, acting as a barrier against
 detection of faint background sources.  }
This barrier 
is remarkably similar to the detection threshold
proposed by Boiss\'e et al. (1998). 
This motivated us to investigate 
the  relation between extinction and metal
column density   in DLA systems. 

\subsection{Extinction versus $N_\mathrm{Zn}$ in DLA systems }

The relations (\ref{BigA_Nv}),  (\ref{SmallA_Nv}) and (\ref{glambda})
are valid in the ISM of any galaxy, including   DLA systems,
which are \ion{H}{i} regions of high-redshift galaxies. 
Applying these relations  to a DLA system and to the local ISM,  and
adopting $\mathrm{E_v}$=Zn, 
$\mathrm{E_r}$=Fe and $\lambda=\lambda_V$, we obtain  
\begin{equation}
A_{V} = a_{V,i} \, { g_V \over g_{V,i} } ~  
{ (1-f_{\mathrm{Zn},i}) \over f_{\mathrm{Fe},i} } ~
{ f_{\mathrm{Fe}} \over   (1-f_{\mathrm{Zn}}) } ~
{ (\mathrm{Fe/Zn}) ~ \over (\mathrm{Fe/Zn})_i  } ~
 N_{\mathrm{Zn}}  ~, 
 \label{AVdla1}
\end{equation}
where the quantities in the local ISM are indicated with the index $i$
and those in the DLA system are not labelled. 
To take into account the dependence on metallicity, which varies among DLAs,
we introduce the  metallicity
$Z \equiv N^\mathrm{tot}_{\mathrm{Zn}}/N_\mathrm{H}$,  
where 
$N^\mathrm{tot}_{\mathrm{Zn}} = N_{\mathrm{Zn}} /(1-f_{\mathrm{Zn}})$
is the total zinc column density (gas plus dust). In this way we obtain the expression
\begin{equation}
A_{V} \simeq A_\circ ~
{ G } ~
 f_{\mathrm{Fe}}  ~
{ (\mathrm{Fe/Zn}) ~ \, \over (\mathrm{Fe/Zn})_{\sun} } ~
Z ~ N_{\mathrm{H}} ~,
\label{AVdla2}
\end{equation}
where $A_\circ  \equiv
a_{V,i} \, (1-f_{\mathrm{Zn},i})/ f_{\mathrm{Fe},i}$ and
$G=g_V / g_{V,i}$. 
The constant $A_\circ $ is estimated  using
the value $<\!\! a_V \!\!>$ derived in the previous section 
and the mean values of dust fractions of the local ISM sample
($f_{\mathrm{Zn},i}=0.40$ and $f_{\mathrm{Fe},i}=0.94$). 
We obtain $A_\circ \simeq 1.85 \times 10^{-14}$ mag cm$^2$,
with a relative error of $\simeq \pm 0.15$ dex
including the uncertainties of the dust fractions.
The DLA extinction  
will vary according to the Fe/Zn abundance in the medium,
the iron depletion and the properties of the grains.  
We briefly discuss each one of these  factors.      

Relative abundances of metals   do not
show strong variations in DLAs, 
despite significant changes of the absolute metallicity  $Z$. 
The Fe/Zn ratio  
is  very close to solar in metal-poor stars
with  metallicities typical of DLA systems
(Mishenina et al. 2002; Gratton et al. 2003; Nissen et al. 2004;
see discussion in Vladilo 2004).  
Deviations from the solar ratio are measured in stars with 
extremely low metallicities. Such deviations  do not affect
the present work in any case, because 
the dust fraction and the extinction vanish
 at very low metallicity,   as we now discuss. 
 We adopt therefore
${ (\mathrm{Fe/Zn}) / (\mathrm{Fe/Zn})_{\sun}  } = 1$.

The depletion of DLAs increases with metallicity (Ledoux et al. 2003). 
In particular,  the dust fraction of iron  undergoes a fast rise between 
between metallicities [Zn/H] $\sim -2$ dex  and $-1$ dex, from 
values close to zero up to  values typical 
of the Milky-Way warm interstellar gas
 (Vladilo 2004). 
Here we model this trend with  the analytical expression
$$
f_{\mathrm{Fe}}(Z) = 
{ 1 \over \pi}  \,
\left[ 
\arctan
\left({\mathrm{[Zn/H]} - \mathrm{[Zn/H]}_\circ \over 
\Delta(\mathrm{[Zn/H]})} \right)+{\pi \over 2} 
\right]   
$$
designed to smoothly fit   the ISM  
and   DLA data, providing  a gradual decrease to zero 
at very low metallicity. 

The dependence on the changes of the grain properties enters through
the factor $G$, which scales  as
$s \, Q_{e,\lambda} \, (r_\mathrm{gr} \, \varrho_\mathrm{gr} \, X^{d}_\mathrm{Fe}) ^ {-1}$
(Eq. \ref{glambda}). 
This term is mostly determined by the physical
conditions of the medium, rather than by its chemical enrichment,
with the possible 
exception of the abundance by mass of iron in the grains, $X^{d}_\mathrm{Fe}$. 
At the very early stages of chemical evolution we may expect variations of
$X^{d}_\mathrm{Fe}$ as a consequence of non-solar relative abundances
of metals. However, these effects do not affect our estimate of the
extinction since, at the very low metallicities typical of these early stages
($Z \ll Z_{\sun}$), the factor $f_{\mathrm{Fe}}(Z)$
vanishes in any case, yielding a null extinction. 

The variations of $G$ due to changes of the physical conditions can be
assessed by sampling different regions     in a medium of constant
composition, such as the local ISM. 
The low scatter of $a_V$ derived in the previous section
indicates that the scatter of $G$ in the local ISM   is  low,
even if the MW sample   includes   lines of sight 
with   different physical conditions.   
To explain this qualitatively, we can imagine that variations of the  
grain size $r_\mathrm{gr}$ 
may compensate variations of   the dust density $\varrho_\mathrm{gr}$.
For instance,  in a harsh interstellar environment, volatile elements
tend to leave the grains and we may expect an increase of
$\varrho_\mathrm{gr}$ because carbon, the element potentially    
most abundant in the grains,  is volatile and has  
low atomic mass. At the same time,  
we may expect that larger grains will be more easily destroyed
in a   harsh environment
and the mean      $r_\mathrm{gr}$ will decrease,
countering the increase of $\varrho_\mathrm{gr}$ in Eq. (\ref{glambda}).
%

These general considerations apply to any type of interstellar environment, suggesting that $G$ may not vary dramatically in DLAs with moderately
low abundances ($Z \ga 0.1 \, Z_{\sun}$). To   test
this hypothesis we computed $G$ in the Small Magellanic Cloud (SMC)
 from the expression
\begin{equation}
{ G_{\mathrm{SMC}}  } \simeq
{ ( { A_V \over N_\mathrm{H} } )_\mathrm{SMC} \over 
   ( { A_V \over N_\mathrm{H} } )_\mathrm{MW}   } \, 
\left[ 
{ Z_\mathrm{SMC}  \over Z_{\sun} } \,
{ f_{\mathrm{Fe},\mathrm{SMC}} \over f_{\mathrm{Fe},\mathrm{MW}} } 
\right]^{-1} \,
\end{equation}
obtained by applying Eq. (\ref{AVdla2}) to the SMC with
$ (\mathrm{Fe/Zn})_\mathrm{SMC} = (\mathrm{Fe/Zn})_{\sun}$.
Inserting
$( { A_V \over N_\mathrm{H} } )_\mathrm{SMC} = 8.7 \times 10^{-23}$ mag cm$^{2}$
(Gordon 2003; SMC bar), 
$( { A_V \over N_\mathrm{H} } )_\mathrm{MW} = 2.1 \times 10^{-22}$ mag cm$^{2}$
(Bohlin et al. 1978),
$ Z_\mathrm{SMC}  \simeq 0.25 \, Z_{\sun} $
(Russel \& Dopita 1992), and
$ f_{\mathrm{Fe},\mathrm{SMC}} (Z_\mathrm{SMC})
\simeq 0.91 \, f_{\mathrm{Fe},\mathrm{MW}}$
 from 
 our analytical relation $f_{\mathrm{Fe}}=f_{\mathrm{Fe}}(Z)$, 
 %
 we obtain
${ G_{\mathrm{SMC}} } \simeq 0.6 $. 
Considering the uncertainties in this derivation, this result is
  consistent with the Milky-Way value, 
suggesting that, indeed, the term $G$ may not vary
dramatically in galaxies of moderately low abundances
even if the extinction curve is different.  
%

Eq. (\ref{AVdla2}) is valid in the rest frame of the \ion{H}{i} region.
To derive the extinction in the observer's frame
we   need to consider  the wavelength dependence of
the extinction.  
In the rest frame of the DLA system
the extinction 
at the wavelength $\lambda^\prime$  will be 
$ A_{\lambda^\prime} \simeq \xi(\lambda^\prime)  \,\, A_{V} $,
where $\xi(\lambda^\prime)=A(\lambda^\prime)/A(\lambda_V)$
is, by definition,  the normalized extinction curve. 
In the observer's frame
the same extinction 
will appear  at   $\lambda = \lambda^\prime \, (1+z)$, 
where $z$ is the  absorption redshift.

\begin{figure}
   \centering 
    \includegraphics[width=7.5cm,angle=0]{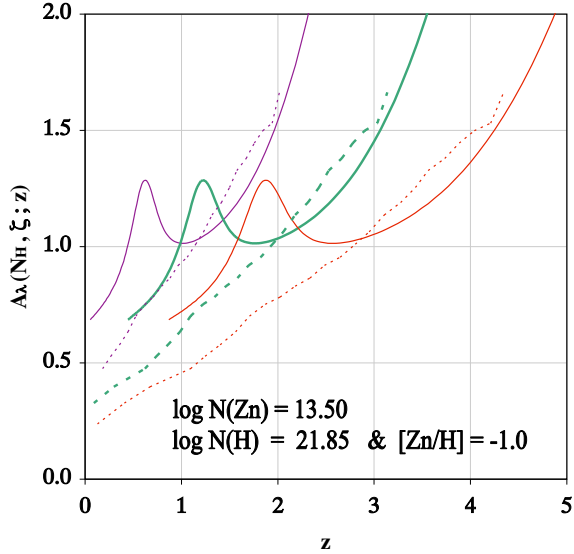} 
\caption{ 
Extinction of a quasar at a fixed wavelength
in the observer's frame, plotted versus
the redshift of an intervening  DLA system with
$N^\mathrm{tot}_\mathrm{Zn} = 10^{13.50}$ cm$^{-2}$. 
Solid and dashed curves represent MW-type and SMC-type
extinction curves, respectively (Cardelli et al. 1998; Gordon et al. 2003). 
Curves left to right: photometric bands $u^\prime$, $g^\prime$, and $r^\prime$, respectively. 
 }
\label{ext_z}
\end{figure}

In conclusion, the observer's frame extinction of a quasar 
at wavelength $\lambda$ due to an intervening DLA system 
at redshift $z$ is
\begin{equation}
A_{\lambda} (N_\mathrm{H},Z; z) 
\simeq A_\circ ~ G ~\,
\xi \! \left(  \! { \lambda \over 1 \!+ \! z } \! \right)  ~
f_{\mathrm{Fe}}(Z) ~\,
 N_{\mathrm{H}} ~ Z ~ ,
\label{AlamHzeta}
\end{equation}
with  
$A_\circ \simeq 1.85 \times 10^{-14}$ mag cm$^2$, 
and $G=1$ and $0.6$ 
for Galactic-type and SMC-type dust, respectively. 
For each type of dust we adopt the corresponding extinction curve
representative of the Milky Way and of the SMC. 


Since $\xi(\lambda^\prime)>1$ for $\lambda^\prime < \lambda_V$,
the net effect of the cosmological redshift is an
amplification of  the rest-frame extinction
at $\lambda <  \lambda_V ~ (1+z)$.
At $z \ga 2$ this amplification compensates for the reduction of  
$f_{\mathrm{Fe}}(Z)$ in systems of moderately
low metallicity and for the mild reduction of $G$ in SMC-type dust. 
Only at very low metallicity the extinction becomes negligible
 since $f_{\mathrm{Fe}}(Z)$ tends to vanish. 
Overall, 
{\em the extinction per unit column density of zinc atoms
predicted for most DLAs
 is similar, or even larger,
than that typical of the Milky-Way ISM}. 
%
%

Examples of quasar extinction   
estimated with Eq. (\ref{AlamHzeta})  
are plotted in Fig. \ref{ext_z} as a function of $z$, keeping
fixed the other terms. This type of representation puts in evidence
the amplification of the extinction with increasing redshift. 
Each curve was computed at a constant $\lambda$  equal to the effective
wavelength of  the SLOAN Digital Sky Survey (SDSS)
photometric bands $u^\prime$, $g^\prime$ and $r^\prime$  
(Fukugita et al. 1996). 
We adopted the extinction law by Cardelli et al. (1998;
hereafter CCM law) with $R_V=3.1$
for the Milky-Way extinction curve 
and the mean SMC bar data  by Gordon et al. (2003) for the  SMC curve.
The term $f_\mathrm{Fe}(Z)$ was estimated at  
 [Zn/H]\,$= \log (Z/Z_{\sun}) = -1$ dex.
The   total zinc column density was fixed
at $N^\mathrm{tot}_{\mathrm{Zn}} = N_\mathrm{H} \, Z
=10^{13.50}$ atoms cm$^{-2}$,
 roughly two times the zinc column density 
threshold proposed by Boiss\'e et al. (1998).
For the   redshift interval typical of most DLAs ($z \approx 2/3$)
the predicted extinction 
is $\gsim 1$ mag, a value sufficient to obscure quasars 
in  spectroscopic surveys. 

 \begin{figure}
   \centering
 \includegraphics[width=7.5cm,angle=0]{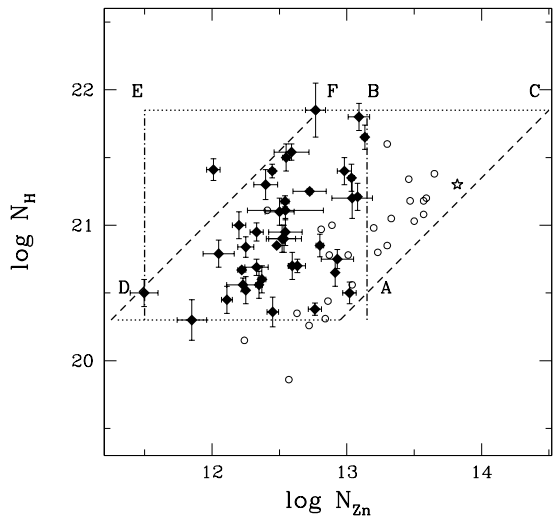}
\caption{Diamonds: 
column densities $N_\mathrm{H}$  versus $N_\mathrm{Zn}$
in DLAs. 
Horizontal lines:  DLA definition threshold  (bottom) and empirical upper
boundary (top) of $N_\mathrm{H}$. 
Vertical lines: detection limit (left) and obscuration  threshold (right) of $N_\mathrm{Zn}$. 
Dashed lines: lines of constant metallicity.
Circles: Milky Way data of Fig. 1. 
Star: GRB-DLA from Savaglio et al. (2003). 
}
\label{sample}
\end{figure}

\section{Evidence for extinction from DLA data}

To search for some extinction effect related to the zinc column density
%
we  collected  from the literature all   DLAs  with 
available $N_\mathrm{Zn}$ measurements. 
The resulting sample, shown in Fig. \ref{sample},
 includes 41 measurements, most of which derived from
 Hires/Keck and UVES/VLT  spectra. 
 References to these data can be found in Vladilo (2004; table 1), with
 the exception of 4 systems (without \ion{Fe}{ii} lines)
  published by Pettini et al. (1994, 1997) and
 Boiss\'e et al. (1998). 
%
 %

The updated sample includes more than twice  the measurements
of the one discussed by Boiss\'e et al. (1998) but,
in spite of this increase,  
column densities  above the original
threshold 
(line AB in the figure) are not   found. 
%
%
The lack of DLAs with $N_\mathrm{Zn} > 10^{13.1}$ atoms cm$^{-2}$
is surprising, since  Milky Way lines of sight with
\ion{Zn}{ii} measurements do show values above this limit
in a significant fraction of cases 
(Fig. \ref{sample}). 
Studies of gamma-ray bursts (GRBs) observed
inmediately after the explosion demonstrate that zinc column densities
above the threshold exist and can be detected in extragalactic absorbers\footnote{
Absorbers detected in front of GRBs, however, may have 
different properties from classical DLAs; 
a claim for gray extinction in GRB/DLAs has been made by
Savaglio et al. (2003). 
}
(Savaglio et al. 2003;
the only case with measured $N_\mathrm{H}$ is indicated
with a star in Fig. \ref{sample}). 

These results confirm that  some selection effect is preventing the  
detection of
values of $N_\mathrm{Zn}$ above the threshold 
when the background source is faint. 
%
%
%
The remarkable similarity between the DLA cutoff at 
$N_\mathrm{Zn} = 10^{13.1}$ atoms cm$^{-2}$  
and  the ISM turning point  
at  $N_\mathrm{Zn} \simeq 10^{13.2}$
 suggests that the rapid rise of the extinction  
above the turning point  can be  responsible for the cutoff. 
This is consistent with the conclusion of the previous section
that  the extinction per unit zinc column density of many DLAs
is similar to that of the local ISM. 

%

\begin{figure}
   \centering
 \includegraphics[width=7.5cm,angle=0]{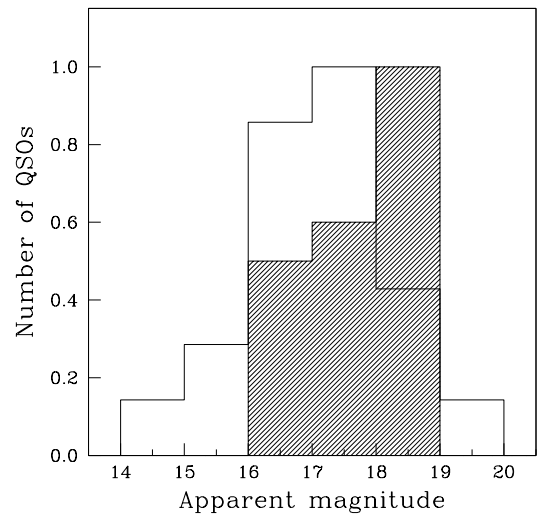}
\caption{Normalized frequency distribution  of quasar apparent magnitudes 
for the 
 sub-sample of
 DLAs with $\log N_\mathrm{Zn} > 12.53$ (shaded histogram)
 and $\log N_\mathrm{Zn} \leq 12.53$ (light histogram).}
\label{Vmaghist}
\end{figure}

If the     extinction increases with $N_\mathrm{Zn}$, we   
expect to find evidence of this effect also below the threshold,
in the range    
$\log N_\mathrm{Zn} \la 13.1$
where we do detect DLAs.
To search for this evidence we analysed the frequency distribution
of quasar apparent magnitudes of the zinc sample. 
We used the $V$ magnitude, which is measured in most cases.
In 4 quasars without $V$ measurements we adopted the magnitude
of nearby optical bands. 
In practice, we compared the behaviour of the two sub-samples
with  $\log N_\mathrm{Zn}$ below and above the median
value, $(\log N_\mathrm{Zn})_\mathrm{median}= 12.53$. 
%
The differences between the two distributions, shown in Fig. \ref{Vmaghist}, are consistent
with a rise of the quasar extinction with increasing $N_\mathrm{Zn}$.
The analysis  of the three bins with more statistics 
 indicates that the maximum of the sub-sample with high $N_\mathrm{Zn}$ (shaded histogram) is 
shifted  by $\approx 1$ bin
relative to the maximum of the other sub-sample (light histogram). 
 A Kolmogorov-Smirnov test shows that there is only a 9.7\% level of
probability that the two distributions are drawn from the same parent
population.
The shift to fainter magnitudes suggests that even at $\log
N_\mathrm{Zn} \lsim 13. 1$ dust extinction is already present,
affecting more the high-$N_\mathrm{Zn}$ sub-sample. 
The magnitude of this effect can only be explained by invoking
a redshift amplification of the extinction, 
as predicted by  relation (\ref{AlamHzeta}), since the extinction
expected at rest-frame is  only $\simeq 0.35$ mag 
at $\log N_\mathrm{Zn} = 13. 1$ (Fig. \ref{figAV_NZn}). 

\section{ Mathematical formulation of the bias }

In   Eq. (\ref{AlamHzeta})
the variation of the depletion in DLAs is accounted for
by the term $f_\mathrm{Fe}(Z)$, which is determined by the metallicity.
The term $G$ is approximately constant, at least in galaxies
with metallicity similar to those of the SMC and Milky Way.
Possible variations of $G$
in galaxies of very low metallicity do not affect the prediction of the extinction
since, in any case, $f_\mathrm{Fe}(Z)$ vanishes when $Z \ll Z_{\sun}$.   
For these reasons, relation (\ref{AlamHzeta}) allows us to estimate 
the extinction of a DLA system using, in practice, only its \ion{H}{i} column density
and metallicity. 
%
%
We use this property to derive a mathematical formulation  
of the obscuration bias  based on the study of the 
 distribution functions of
$N_\mathrm{H}$ and $Z$ in DLA systems. 
%
In practice, we determine the fraction of DLAs  
that are missed as a consequence of their own extinction.
We then assume that multiple DLA absorbers in a given line of sight
play a negligible role in the estimate of the obscuration bias. 
%
In this way,  as we show in the Appendix, we   derive
a   mathematical formulation of the   bias
which only makes use of observable statistical distributions
of DLAs and quasars.
%
To quantify the obscuration effect we need   a good statistics
of the observed distributions and this requirement limits
the redshift range where  our method can be presently applied,
since the bulk of DLAs data are currently concentrated
in $1.8 \lsim z \lsim 3.0$. 

An advantage of our formulation 
is that it does not require a knowledge of 
the intrinsic luminosity function of quasars or
 the geometrical distribution of the neutral gas in DLA galaxies. 
We refer to   FP93 for a   mathematical
treatment of the effect of quasar obscuration
%
 which describes the relation
with the quasar luminosity function and the geometrical distribution of the gas
in the intervening galaxies.   

\subsection{ True and biased distributions }

 %
We   consider the DLAs in  the redshift interval 
$(z_\mathrm{min},z_\mathrm{max})$
that are detectable in a survey with limiting magnitude $m_\ell$.
We call
$f_{N_\mathrm{H}} d N_\mathrm{H}$   the  
  number of  such  DLAs with  column densities between
$N_\mathrm{H}$ and  $N_\mathrm{H}+dN_\mathrm{H}$  
and $f_{Z} \, dZ$ the    number of   those with
metallicity between $Z$ and $Z+d \,Z$.
The distributions $f_{N_\mathrm{H}}$ and $f_{Z}$ are defined in absence
of obscuration bias and we call them  the "true" distributions.
When we consider the effect of the  
 extinction generated by     these DLAs, 
the number of detectable systems
  will be  
$f^\mathrm{b}_{N_\mathrm{H}} d N_\mathrm{H}$ 
and
$f^\mathrm{b}_{Z} \, dZ$.
We call $f^\mathrm{b}_{N_\mathrm{H}}$ and $f^\mathrm{b}_{Z}$ the "biased" distributions. 
In the Appendix we derive the following relations between true and biased
distributions  
[see Eqs.  (\ref{fb_H}) and (\ref{fb_zeta})],
 \begin{equation}
 f^\mathrm{b}_{N_\mathrm{H}}  \simeq 
{
\int_{0}^{\infty}
  \mathcal{B}_{m_\ell}  ( N_\mathrm{H}, Z) \,
f_{Z}  \, d\, Z
 \over 
\int_{0}^{\infty}
f_{Z}  \, d\, Z
 } ~  f_{N_\mathrm{H}} ~~
 \label{fb_H_paper}
\end{equation}
and 
%
%
\begin{equation}
 f^\mathrm{b}_{Z}  \simeq   
{
\int_{N_\mathrm{DLA}}^{\infty} 
  \mathcal{B}_{m_\ell}  ( N_\mathrm{H}, Z) \,
f_{N_\mathrm{H}}  \, d\, N_\mathrm{H} 
 \over 
\int_{N_\mathrm{DLA}}^{\infty} 
f_{N_\mathrm{H} }  \, d\, N_\mathrm{H}
 } ~  f_{Z} ~~.
 \label{fb_zeta_paper}
\end{equation}

For simplicity  we omit  the dependence of the equations
on $z_\mathrm{min}$ and $z_\mathrm{max}$.
We assume that
   the statistical distributions  of interest       smoothly vary
inside the redshift interval, so that we can approximate them   
with their value at mean redshift  $\overline{z}=(z_\mathrm{max}-z_\mathrm{min})/2$,
  where     
$\overline{z}$   
  is  close to  the peak of the observed distribution.

The "bias function"
$
\mathcal{B}_{m_\ell} (N_\mathrm{H}, Z) 
\equiv 
B_{m_\ell} (N_\mathrm{H}, Z, \overline{z}) 
$
that transforms the   true distributions      
into the biased ones is given by the relation
[see Eq. (\ref{BigBHZz})]
\begin{equation}
B_{m_\ell}  ( N_\mathrm{H}, Z,z) \equiv
{
  \int_\circ^{m_{\ell}-A_\lambda(N_\mathrm{H},Z, z)} \,  
 n(m; z)  \, d m \,
  \over
  \int_\circ^{m_\ell}  \, n(m; z)  \,  dm \, ,
 \label{BigBHZz_paper}
 } 
\end{equation}
where
$A_\lambda(N_\mathrm{H},Z, z)$ is the DLA extinction
and $n(m; z)$ the quasar magnitude distribution. 

 In the next section we show how to estimate $n(m; z)$, while
 in Section 4.3  we present a procedure
 for deriving the true distributions from Eqs.  (\ref{fb_H_paper}) and (\ref{fb_zeta_paper}),
 starting from the observed distributions. 

In deriving these equations
we made the assumption that the
distributions of $N_\mathrm{H}$     and $Z$
are statistically independent. We discuss this assumption in Section 4.4.
%

\subsubsection{ The quasar magnitude distribution }
 
We call  $n(m; z) \, dm$   the
  number  of quasars  with apparent magnitude  $m \in (m,m+dm)$
observable beyond   redshift   $z$  all over the sky.
This  number is defined 
in absence of quasar obscuration and we call
 $n(m; z) \, dm$  the "true" distribution.
 %
%
Our goal is to infer the true   distribution from the observed one, 
which must equal
the distribution  biased by the obscuration effect, $n^\mathrm{b}(m; z)$.
The  relation between true and biased distribution is discussed 
 in the Appendix (Section A.2),
where we derive the equations Eq. (\ref{nb_m}), i.e.
 \begin{equation}
n^\mathrm{b}(m; z)
\simeq 
\left[ 1-\overline{\mathcal{F}}_1 \right] \, n(m; z) 
+ \, \overline{\mathcal{F}}_1 \, n^\mathrm{b}_1(m;z) ~,
 \label{nb_m_paper}
\end{equation}
and Eq. (\ref{p_m}), i.e.
\begin{equation}
n^\mathrm{b}_1(m; z) = K 
\int_{-\infty}^{\infty} 
n(m^\prime; z) ~ 
f_{A_\lambda}\!(m-m^\prime;  \overline{z}_e) ~ dm^\prime ~.
\label{p_m_paper}
\end{equation}
In these relations 
$\overline{\mathcal{F}}_1$ is 
the fraction of quasars with one foreground DLA,
$n^\mathrm{b}_1(m; z) $   the magnitude distribution
of quasars with one foreground DLA, 
  $K$ is a normalization factor and
$f_{A_\lambda}(A_\lambda; \overline{z}_e)$ 
  the frequency
distribution of the extinctions 
of   the DLAs in the redshift range $z \in (0,\overline{z}_e)$
($\overline{z}_e$
is the mean redshift of the quasars in the survey). 
In Section 4.3, we show how to   use Eqs. (\ref{nb_m_paper}) and (\ref{p_m_paper}) 
to infer  $n(m; z)$ (see also Appendix A.2).


 \subsection{Obscuration fraction}
 
 We define the obscuration fraction  using the mathematical relations
 between the true and biased distributions. 
 We call {\em total obscuration fraction} 
 the quantity
 \begin{equation}
\Phi_{m_\ell} 
= 1 - 
{
\int^\infty_{N_\mathrm{DLA}}  f^\mathrm{b}_{N_\mathrm{H}} \, d N_\mathrm{H} 
\over 
\int^\infty_{N_\mathrm{DLA}}  f_{N_\mathrm{H}} \, d N_\mathrm{H} 
}
~.
\label{BigPhi}
\end{equation}
The total obscuration fraction
  represents the   number   ratio
of   DLAs missed due to quasar obscuration
in a survey with limiting magnitude $m_\ell$. 
The dependence on the limiting magnitude enters in
$f^\mathrm{b}_{N_\mathrm{H}} $ through Eqs. (\ref{BigBHZz}) and (\ref{fb_H}).
The total obscuration fraction can also be derived 
using the true and biased distributions of metallicity. 
%

We call  {\em  obscuration fraction}, $\phi_{m_\ell}$, 
the fraction by number of  systems obscured at a particular value     
of column density or metallicity.
 For instance, the fraction of systems obscured at a specific
  value of $N_\mathrm{H}$ is
 \begin{equation}
 \phi_{m_\ell} (N_\mathrm{H}) = 1 - 
 {f^\mathrm{b}_{N_\mathrm{H}}  \over f_{N_\mathrm{H}}} 
 ~~.
 \label{phi}
 \end{equation}
Similar expressions can be derived 
to define the  obscuration fractions 
$ \phi_{m_\ell} (Z)$ and $ \phi_\ell (N^\mathrm{t}_\mathrm{Zn})$, 
i.e. the fraction of systems missed at a
particular value of $Z$ and $N^\mathrm{t}_\mathrm{Zn}$.

\subsection{ Solution of the equations  }

%
The true   functions 
$f_{N_\mathrm{H}}$ and $f_{Z}$ are the unknown
quantities in Eqs. (\ref{fb_H_paper}) and (\ref{fb_zeta_paper}). 
The procedure that we follow to solve these equations
consists in 
(i) adopting trial functions $f_{N_\mathrm{H}}$ and $f_{Z}$,
(ii) estimating  $n_m \equiv n(m; \overline{z})$,
(iii) computing $f^\mathrm{b}_{N_\mathrm{H}}$ and $f^\mathrm{b}_{Z}$
from  the equations
and (iv) comparing these
biased distributions with the observed ones.
These steps are repeated by changing the trial functions
until the predicted biased functions match the observed ones.  
%
%
We briefly describe each step of this procedure.

In the first step, 
we adopt analytical functions in parametrical form
as the trial functions $f_{N_\mathrm{H}}$ and $f_{Z}$.
%
%
%
%
Once we choose   the   shape, the problem
of determining the true functions $f_{N_\mathrm{H}}$ and $f_{Z}$ 
is equivalent to that of determining the values of their parameters.
%
 
In the second step we use the trial  functions
 $f_{N_\mathrm{H}}$ and $f_{Z}$
to derive the  distribution of extinctions
from Eq. (\ref{AlamHzeta}).
We use this extinction distribution
in Eq. (\ref{p_m_paper}). 
We then iterate Eqs. (\ref{nb_m_paper}) and  (\ref{p_m_paper}),
starting with a trial function\footnote{
The observed distribution  $n^\mathrm{b}(m; z)$ can be used
as the starting trial function of $n(m; z)$.}  
  $n(m; z)$, until we find  a $n(m;z)$
  that yields a perfect agreement between
$n^\mathrm{b}(m; z)$ and  the  
  distribution $n^\mathrm{obs}(m; z)$
measured from the surveys. 
%

In the third step we  use $n(m; \overline{z})$ to derive    
$\mathcal{B}_{m_\ell}  ( N_\mathrm{H}, Z)$
from  Eq. (\ref{BigBHZz_paper}) 
and, finally, the biased 
distributions  
$f^\mathrm{b}_{N_\mathrm{H}}$ and $f^\mathrm{b}_{Z}$ 
from Eqs.  (\ref{fb_H_paper}) and (\ref{fb_zeta_paper}). 

In the last step  we compute the $\chi^2$ deviation between
the predicted distributions
$f^\mathrm{b}_{N_\mathrm{H}}$ and $f^\mathrm{b}_{Z}$ 
and the corresponding observed distributions.

By iterating the steps we determine
the  parameters of $f_{N_\mathrm{H}}$ and $f_{Z}$
which allow     
$f^\mathrm{b}_{N_\mathrm{H}}$ and $f^\mathrm{b}_{Z}$ 
to best fit the empirical distributions.  
The search  is done via $\chi^2$ minimization. 
Since we are not guaranteed a priori of the unicity of the solution,
we carefully investigate the variation of $\chi^2$ in the parameter space
to make sure the minimum is unique.  

\subsection{Are $N_\mathrm{H}$ and
 $Z$ statistically independent   ?}

The mathematical formulation presented above
relies on the assumption that $N_\mathrm{H}$ and
 $Z$  are statistically independent of each other in the true population of DLAs.  
The local ISM offers an example in support of this independence, 
since a broad range of \ion{H}{i} column densities,
including the high values typical of DLAs,   are found at   constant  
metallicity, $Z \simeq Z_{\sun}$.
Similarly,  a DLA system of given metallicity $Z \simeq Z_\circ$
will display a distribution of  \ion{H}{i} column densities, if
sampled along random lines of sight. 
This  distribution will reflect the mass spectrum of interstellar clouds,
which is likely to be determined by some physical mechanism,  
such as interstellar turbulence (see discussion in  Khersonsky \&
Turnshek 1996 and refs. therein), rather than by the  metallicity.  
In this sense, it is reasonable to assume that   
  $f_{N_\mathrm{H}}$ and $f_{Z}$  are independent distributions.
  
  Independent support of these   arguments comes
from the    
hydrodynamic simulations of DLAs evolving in $\Lambda$CDM
cosmology (Cen et al. 2003), 
which indicate that metallicity and hydrogen 
column density are almost completely uncorrelated.  
Similar results are found from the  cosmological SPH simulations
performed by Nagamine et al. (2004).

Notwithstanding, a trend is expected because the metallicity  
correlates with the star formation rate, 
which in turn increases with the gas density in the host galaxy. 
Therefore, lines of sight with high $N_\mathrm{H}$ may have a
tendency to be associated with higher metallicities.
A   weak evidence for this trend was found
by Cen et al. (2003), but
the effect is particularly small at redshift $z \simeq 2-3$.  
 If the metallicity is correlated with $N_\mathrm{H}$, 
the effect of the obscuration is   more important than 
in the ideal case considered above. In this case
  the application of our mathematical formulation 
  gives a conservative estimate of the bias. 

\subsection{Gravitational magnification}

Competing with the obscuration effect, is the possible gravitational magnification 
of the background quasar by the absorber itself. 
Many studies over the years have attempted to estimate
the magnitude of this phenomenon for both DLAs (Smette et al. 1997, Le
Brun et al. 2000) and metal absorption lines (Vanden
Berk et al. 1996, M\'enard \& P\'eroux 2003). The results show that for the DLAs
selected at optical wavelengths, gravitational effects are small.
%
%
In the present application of our method, 
based on DLAs selected at optical wavelength, we therefore neglect  
the magnification effect.
%

\section{Implementation of the procedure}

We   describe the statistical distribution functions   adopted here
as a first example of   implementation of the  procedure. 
For the intrinsic distributions we only need to adopt
a functional dependence (i.e. a shape),  
leaving to the procedure the task of
determining the parameter values.

\subsection{ The empirical distributions}

 We consider the sub-sample of
DLA systems with available measurements of \ion{Zn}{ii} column densities,
for which it is possible to derive metallicities corrected
for depletion effects   (Vladilo 2004). 
The differences between the corrected metallicities and
the [Zn/H] values taken at face value are small, with a mean value
 of $+0.12$ dex.  
%
Given the size  of the zinc sub-sample ($\sim 40$ systems;
see list of references in Table 1 of Vladilo 2004) we can estimate 
the frequency distribution of metallicities 
 with a Poisson statistical error $\la 10\%$
 in only a few bins.
To make the best use of the available data we  selected  
the systems in the redshift range $1.8 \leq z \leq 3$,
where most of the $N_\mathrm{Zn}$ measurements have
been obtained so far.  
The mean redshift of the sample (28 systems)
is $z \simeq 2.3$.
%
 %
To derive the empirical distribution 
of $N_\mathrm{H}$ we used the surveys by
Storrie-Lombardi \& Wolfe (2000) and
P\'eroux et al. (2003),
which have more statistics  than the zinc sub-sample. 
Selecting the DLAs in the range $1.8 \leq z \leq 3$ 
we obtained a  sample of $59$ systems.

Then
$\log N(\ion{H}{i})$ and [Zn/H]=$\log (Z/Z_{\sun})$, 
we transformed the derived
distributions in linear space, $N(\ion{H}{i})$ and $Z$,
for later comparison with the distributions $f_\mathrm{H}$ and $f_Z$.
In computing  the  biased distributions
with our procedure, we   took into account the fact that
the typical  limiting magnitude of the  $N_\mathrm{H}$ surveys
($m_\ell \sim 19.5$)
is somewhat larger than that of the  
metallicity surveys ($m_\ell \sim 19$), owing to the different requirements
in spectral resolution.  
In fact, Eqs. (\ref{fb_H}) and (\ref{fb_zeta}) 
can be solved independently of each other, inserting in each
case the appropriate value of $m_\ell$.

\subsubsection{The frequency distribution of quasar magnitudes}

\begin{figure}
  \centering
 \includegraphics[width=7.cm,angle=0]{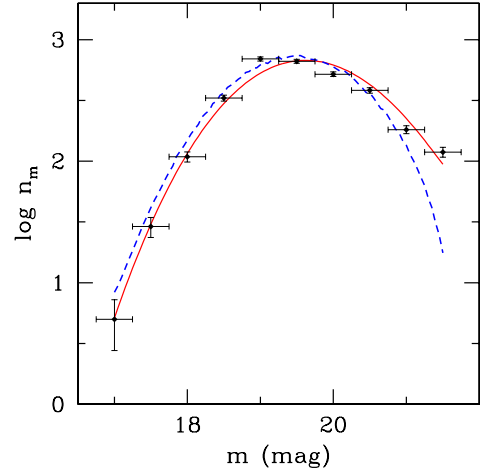}
\caption{ 
Diamonds: 
distribution of the SDSS $g^\prime$
  apparent magnitudes  for
quasars in the   interval $2 \leq z_e \leq 5$. 
Smooth lines: example of a
pair of model distributions       in which
the "true" model (dashed line) is   determined
from the biased model 
that fits the observations (solid line),
once  the  distribution of DLA extinctions  
and $\overline{\mathcal{F}}_1$ are specified
(see Section 4.1.1). 
}
 \label{plotQSOmag}
\end{figure}

To derive   the empirical distribution of quasar magnitudes
we used the data of the Sloan Digital Sky Survey (SDSS)  (Schneider et al. 2003).
The distribution was determined for two photometric bands  well representative
of the visual part of the spectrum, namely the
$g^\prime$ and $r^\prime$ bands, with effective wavelengths 
$\lambda=0.48 \mu$m and $0.62 \mu$m, respectively
(Fukugita et al. 1996). 
In each band we binned  the data   
at   steps of  0.5 mag,  counting all the quasars located beyond    
the typical absorption redshift of our sample
(in practice, we adopted $ 2 \leq z_e \leq 5 $). 
%
An example of
empirical distribution computed for the $g^\prime$ band 
is shown in Fig. \ref{plotQSOmag} (diamonds with error bars). 
After binning the data, we performed
a polynomial fit to the observed $\log n$ - $\log m$ distribution, up to the 
magnitude $21.5$, for which the statistics   and the
completeness of the SDSS sample are still good. 
%
This fit was adopted as a smooth model of  the observed distribution
in the  magnitude range $ 17 \leq m \leq 21.5$.   
An example of   fit is   shown in Fig.
\ref{plotQSOmag} (solid line).
The limit $m \sim 21.5 $ is sufficient
for studying   the effects of quasar obscuration on the statistics
on DLAs  since  the typical limiting magnitude
of DLA surveys is currently $m_\ell \lsim 21$.

\begin{center} 
\begin{table*}  
\caption{  Results. }
\begin{tabular}{cccccccccc}
\hline \hline
 Ext.curve &  Band & $\beta$ & $\alpha$ & 
$\log ( {Z_\ast \over Z_{\sun} }) $ & 
$\Phi_{22} (m_\ell) ^a $ & $\Phi_{23} (m_\ell) ^b $ &
$\log ( {<Z> \over Z_{\sun}}) $& 
$\Omega_{22}^a$ & $ \Omega_{23}^b$ \\
& & & & & $19.0 / 19.5$ & $19.0 / 19.5$ & & $(10^{-3})$ & $(10^{-3})$ \\
\hline
MW  & $r^\prime$ & 1.56 & $-0.44$ & $-0.13$ & $0.41 / 0.34$ &  $0.44 / 0.38$ & $-0.37$ & $2.3$ & $~7.1$ \\

MW  & $g^\prime$ & 1.51 & $-0.40$ & $-0.10$ & $0.48 / 0.42$ & $0.51 / 0.46$ & $-0.31$ & $2.8$ & $~9.5$\\

SMC & $r^\prime$ & 1.61 & $-0.46$ & $-0.19$ & $0.33 / 0.27$ & $0.37 / 0.30$ &  $-0.44$ & $1.9$ & $~5.5$ \\

SMC & $g^\prime$ & 1.48 & $-0.38$ & $-0.07$ & $0.52 / 0.46$ & $0.56 / 0.51$ & $-0.27$ & $3.1$ & $11.3$\\
\hline
\end{tabular} 
\\
$^a$   
  Upper limit of integration in Eq. (\ref{BigPhi})  or (\ref{Omega})
  truncated at $\log N_\ion{H}{i}=22$. 
\\  
$^b$   
  Upper limit of integration in Eq. (\ref{BigPhi})   or (\ref{Omega})
  truncated at $\log N_\ion{H}{i}=23$. 
\\  
\end{table*}
\end{center}

\subsection{ The shape of the true  $N_\mathrm{H}$ distribution }
 
One way to approximate
 the true shape of   $f_{N_\mathrm{H}}$
is to assume a 
geometrical distribution of the gas   in DLA galaxies. 
For   planar disks with radial  exponential profiles 
$N_{\bot }(r) = N_{\bot 0} \exp (-\alpha \, r)$, 
the expected frequency distribution is
$f_{N_\mathrm{H}} \propto N_\mathrm{H}^{-1} ~ \ln N_\mathrm{H}$
for $N_\mathrm{H}\ll N_{\bot 0}$
and 
$f_{N_\mathrm{H}} \propto N_\mathrm{H}^{-3} $
for $N_\mathrm{H}\gg N_{\bot 0}$
(FP93).
For  central column densities $N_{\bot 0} \ga 3 \times 10^{21}$ cm$^{-2}$
or much larger, as inferred by FP93, this gives  $f_{N_\mathrm{H}} \propto N_\mathrm{H}^{-1} ~ \ln N_\mathrm{H}$
for most of the column-density range typical of DLAs. 

A realistic approximation of $f_{N_\mathrm{H}}$  
should also take into account the fact that, in a given galaxy,
the interstellar clouds show a spectrum of masses, $m_c$, and sizes, $r_c$, 
which is not accounted for by a smooth radial profile. 
A large number of studies indicate that the mass spectrum of 
Milky-Way interstellar clouds follows a power law $f(m_c) \sim m_c^\gamma$,
with $\gamma = -1.5 \pm 0.2$ over a 6 orders of magnitude range
of masses (Scalo \& Lazarian 1996 and refs. therein). 
From this mass spectrum, together with a relation between 
the internal density of the cloud, $n$, and the cloud size 
we can infer a column-density distribution.
Adopting $n \sim r_c^p$ with $-1.2 < p \leq 0$
(Scalo \& Lazarian 1996 and refs. therein) we find\footnote{
With the adopted relation we find   
$m_c \sim N_\mathrm{H}^{p+3 \over p+1}$
and then derive the lower limit of $\beta$ assuming 
$p=0$.
}
that the column-density distribution should follow a power law
$f_{N_\mathrm{H}} \sim N_\mathrm{H}^{-\beta}$ with
$\beta \geq 2.5$. If this column-density spectrum applies to
DLAs, we would expect a decline of column densities faster
than that predicted by the exponential profile model. 

%

Both the exponential profile model and
the cloud mass-spectrum hypothesis suggest that the
true distribution may be approximated with a power law.
 The hydrodynamic computations of DLAs by Cen et al. (2003) 
yield a simulated true distribution of $N_\mathrm{H}$
consistent with  this conclusion.

In light of these considerations, we adopt   
the simple   law 
$f_{N_\mathrm{H}} \sim N_\mathrm{H}^{-\beta}$
for the true distribution, with 
the parameter $\beta$ to be determined by the method itself. 
If this is a good approximation, 
 the resulting biased distribution $f^\mathrm{b}_{N_\mathrm{H}}$
must reproduce the observed fast decline of the number of DLAs
with column density, which 
 is commonly fitted by a   Schechter-type function
\begin{equation}
f_{N_\mathrm{H}} \propto 
{\left( N_\mathrm{H} \over N_\ast  \right)}^{-\beta} \, 
e^{-\left( N_\mathrm{H} \over N_{\ast} \right)} ,~
\label{schechter}
\end{equation}
    characterized by a fast decline at $N_\mathrm{H} > N_{\ast}$
(Pei \& Fall 1995; Storrie-Lombardi \& Wolfe 2000; 
P\'eroux et al. 2003).

\subsection{ The shape of the  true metallicity distribution  }
 
The arguments   leading to the choice of   $f_{Z}$
    can be summarized as follows. 
The observed distribution  of [Zn/H] measurements
peaks at [Zn/H] $\simeq -1.1$ dex and does not show
systems below $\approx -2$ dex
or above solar metallicity  (Pettini et al. 1997).   
Although this distribution is  likely biased at
both ends, we do expect a genuine
decline in the number of DLAs below and
above some critical values of metallicity. 
At   [Zn/H] $\lsim -1.5$ dex  the zinc sample
shows a decrease of the number of DLAs
 in a region of the 
($N_\mathrm{Zn}$, $N_\mathrm{H}$)
plane not affected by   any   bias
(the triangle DEF in Fig. \ref{sample}). 
At high metallicity it is reasonable to predict a decrease since we
expect a natural decline  of the
number of systems with higher and higher star formation rates.  
These arguments indicate that the true  
distribution of [Zn/H] starts from a negligible value
at low metallicity,  shows a rise around $\approx -1.5$ dex
and declines after reaching a maximum.
%
We have been particularly careful in modelling
this trend  with an analytical function
because the results, and in particular the mean metallicity,
depend on the adopted functional form. 
To choose the function we started by making an educated
guess of the shape of the metallicity distribution based
on our knowledge of the statistics of galaxies and DLAs. 
We then made sure that the adopted function satisfies two requirements:
(1)  that it is a "conservative" one, i.e.  a function that does not overpredict
the effects of the obscuration;
(2) that the predicted extension of the function at high metallicities,
where the obscuration is important,
is not affected by the poor knowledge of the low end of the distribution. 
%
After considering several possibilities, 
 we adopted the function
 %
%
%
\begin{equation}
f_Z
 ~ \propto ~ (Z/Z_\ast)^\alpha ~
 e^{-Z/Z_\ast} ~,
 \label{MetalDistFunction}
\end{equation} 
which satisfies very well the above requirements, as we explain below. This function is
similar to the one proposed by Schechter (1976) for
describing the luminosity function of galaxies. 

To make an educated guess for the shape of the   distribution 
we started from  the  metallicity-luminosity relation in galaxies.
Evidence is building up that this  relation is valid not only in the local
Universe  (e.g. Lamareille et al. 2004), but also in DLA systems
(Ledoux et al. 2005). 
The empirical relation   between the logarithmic metallicity 
and  the absolute magnitude
implies that  the linear metallicity $Z$
increases linearly  with the luminosity. The luminosity
in turn follows a Schechter distribution both
in the local Universe (e.g. Cuesta-Bolao \& Serna 2003),
at redshift $z \simeq 0.1$ (e.g. Blanton et al. 2003) and, apparently, up
to   $z \simeq 3.5$ (Poli et al. 2003). 
Combining these different pieces of evidence, it is reasonable
to assume that $Z$ may follow  a   Schechter distribution in DLA systems. 

The     requirement (1) of a "conservative" function   
is equivalent to make sure that         decline at high metallicity 
is fast.  
In fact, the faster the decline of the "true" distribution, the lower
 the estimated number of missed DLAs.
%
Luckily, the Schechter  satisfies this requirement
since it provides a very   fast decline
at $Z \!\!>\!\! Z_\ast$. In fact,
the decline is faster than in       any polynomial in $Z$
and in a lognormal distribution. 

The   requirement (2) implies that we should   avoid functions
for which a single parameter specifies
the behaviour at both ends of the distribution.  %
The Schechter function    satisfies   this requirement
since 
the parameter   $\alpha$, 
which in practice controls the extension of the distribution at low metallicities,
has a negligible effect on the exponential decrease at $Z >  Z_\star$.  
%
%
For a lognormal distribution, instead, 
the width parameter
 is affected 
by the fit   of the distribution at the low-metallicity end,
which is poorly constrained by the observations. 
The resulting error on the width parameter 
will affect the extension of the distribution at high metallicities and therefore
  the estimate of the obscuration effect. 
The same considerations apply to any distributions that,
like the lognormal one, are  specified by a centroid and  a width.
For these types of functions,
the uncertain behaviour of the low end of the metallicity distribution
affects the determination of the obscuration effect.

%


%

\section{Results and discussion}

In Table 1 we summarize the results obtained   from the application
of our procedure.   
The quasar extinctions were   estimated in the
 photometric bands $r^\prime$ and $g^\prime$ 
for an absorption redshift $z=2.3$. 
The differences between the results obtained
 for the two   
 bands give  an estimate of the uncertainty 
due to the lack of a homogeneous set of visual magnitudes 
for all the quasars of the adopted surveys. 
The obscuration fraction   in the $r^\prime$ band is more conservative
than that in the $g^\prime$ band, given the increase
of the extinction with decreasing effective wavelength.
The results for the $g^\prime$ band ($\lambda=0.44 \mu$m) are more
comparable to those derived 
in previous studies of the   obscuration bias,  
which   considered the $B$ band  ($\lambda=0.44 \mu$m;
Ellison et al. 2001).
 
Two types of dust extinction curves were considered, namely the MW
curve by Cardelli et al. (1988) and the SMC curve by Gordon et al. (2003). 
The presence or absence of the 2175\AA\ extinction bump in DLAs
is irrelevant for most of our results, because at $z=2.3$
the extinction curves in the $r^\prime$ and $g^\prime$ bands 
do not sample the bump
 (Fig. \ref{ext_z}).  The only marginal exception is the 
combination of the $r^\prime$ band with the MW-type extinction.
 If the bump is completely absent in DLAs, the MW results
 become closer
to     the SMC results for the $r^\prime$ band. 
 
For each combination of extinction type  and photometric band   
we derived the best-fit values of the parameters 
via   $\chi^2$ minimization
in the intervals 
$1.0 \leq \beta \leq 2.0$, 
$-1.0 \leq \alpha \leq 0.0$, and
$-1.0 \leq \log (Z_\ast/Z_{\sun}) \leq +0.5$.  
The best-fit parameters are listed in Table 1. 
We made sure the minimum is unique
by careful inspection of the variation of $\chi^2$ 
in the specified parameter space. 
The typical fit errors 
of $\beta$, $\alpha$ and $\log (Z_\ast/Z_{\sun})$
are $\simeq \pm 0.12$, $\sim \pm 0.25$  and $\simeq \pm 0.15$ dex,
respectively. 
 %

\begin{figure*}
   \centering
 \includegraphics[width=7.5cm,angle=0]{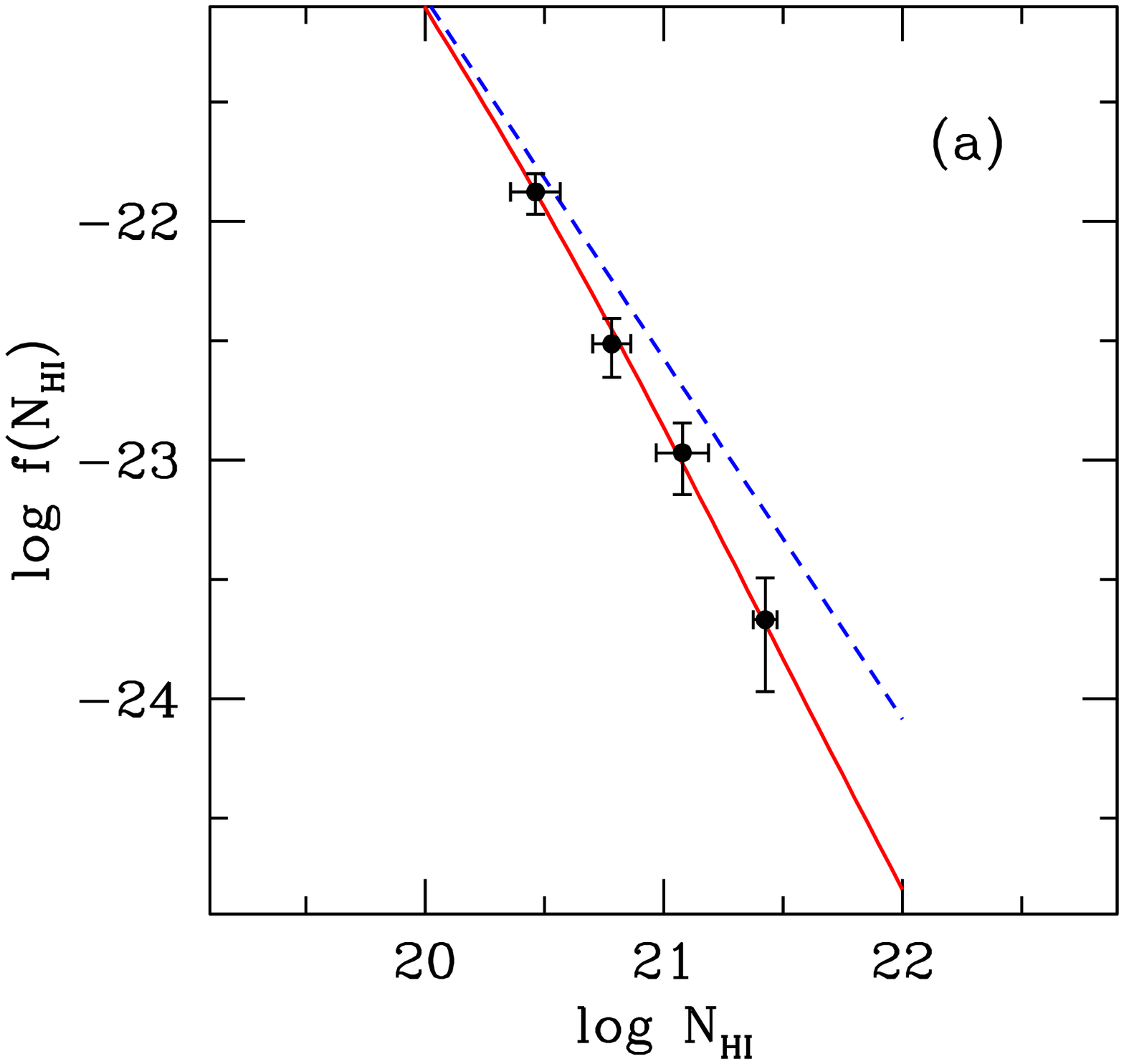} 
 \hspace{0.5cm}
  \includegraphics[width=7.5cm,angle=0]{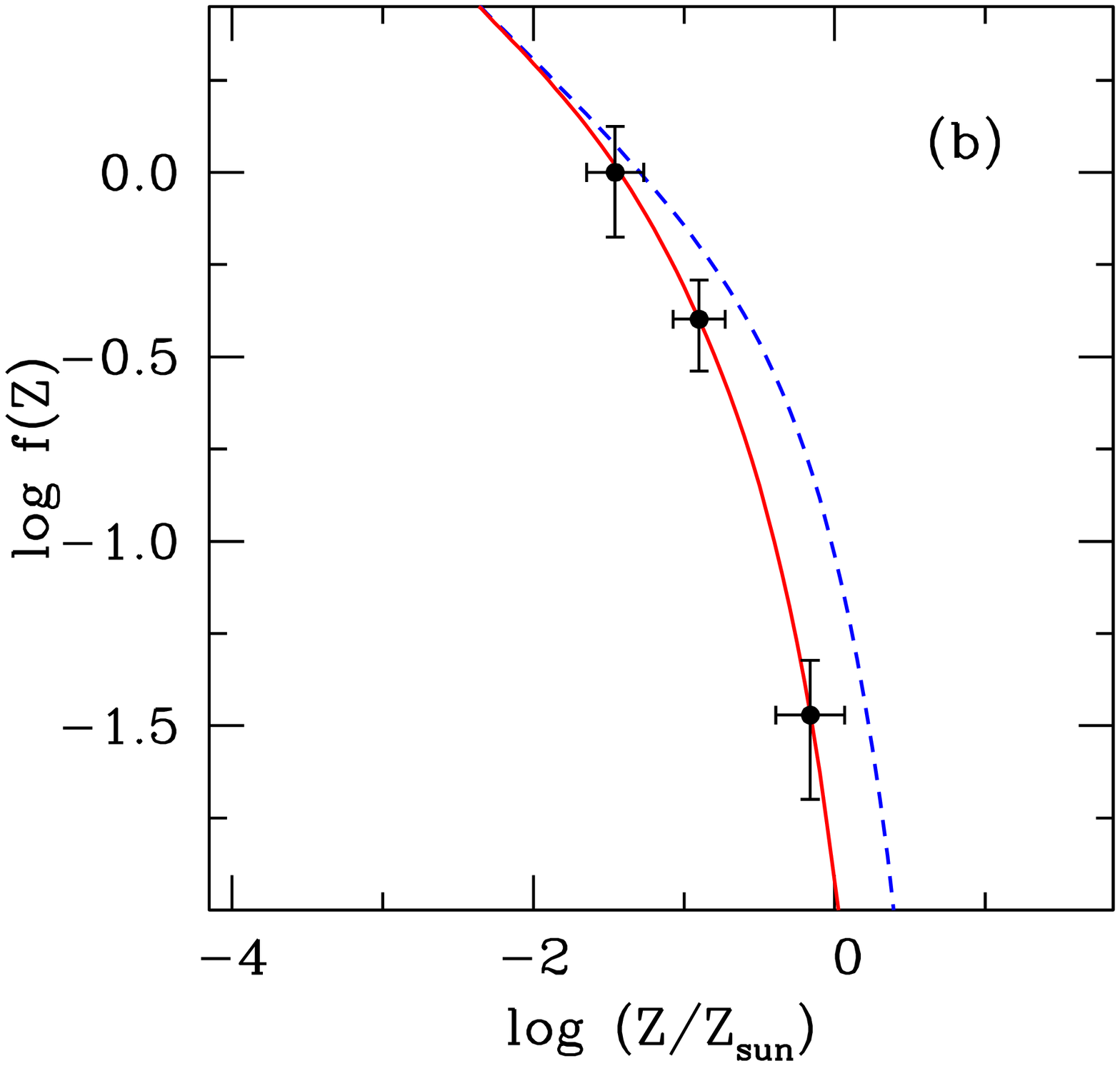} 
  \caption{  Frequency distributions of 
  of \ion{H}{i} column densities (left panel) and metallicities (right panel)
  in DLAs. 
Dashed lines:   true distributions.
Solid lines:  distributions biased by  quasar obscuration.
Circles: empirical distributions obtained from DLA surveys.  
By imposing that the biased distributions best fit the observed ones,
we infer the true distributions (see Section 4.4). The solution shown in the figure was
derived for DLAs with a MW-type extinction and quasars observed
in the $g^\prime$ band.
  }
         \label{fits_mw}%
 \end{figure*}

In Fig. \ref{fits_mw} 
we show the  true and biased distributions (dashed and solid lines) 
of the  best-fit solution  for the   case of
MW-type extinction and $g^\prime$ band. 
The empirical distributions (circles with error bars)
are well reproduced by the computed biased distributions. 
Similar results  are found for all the   cases
considered in Table 1.

The effect of DLA extinction on the shape of
distribution function of quasar apparent magnitudes 
%
can be seen in Fig. \ref{plotQSOmag},
where we show  
the adopted distribution $n_m \equiv n(m;z)$ (dashed line)
and the predicted biased distribution 
$n^\mathrm{b}_m \equiv n^\mathrm{b}(m;z)$ (solid line)
for the $g^\prime$ band and MW-type extinction.     
The agreement between  $n^\mathrm{b}_m$ and
the empirical distribution (diamonds) demonstrates the    
capability of the procedure to account for the bias
in a self-consistent way. 
The results are very stable for variations of $\pm 0.1$
of the parameter
$\overline{\mathcal{F}}_1$ around the adopted value
$\overline{\mathcal{F}}_1 = 0.5$. 
Fig. \ref{plotQSOmag} shows that 
the obscuration effect does not change substantially
the shape of the distribution of apparent magnitude of the quasars, at least
within the current values of limiting magnitude of the surveys.
This is due to the fact that a large fraction of DLAs has   relatively
low \ion{H}{i} column density, as shown in Fig. \ref{fits_mw} and, as a consequence,  
 low extinction. 
The modest bias of the quasar statistics
does not imply that the bias is unimportant
for the DLAs statistics, as we discuss below. 

%
%
Fig.  \ref{fits_mw}{\bf a} is reminiscent of Fig. 8 by FP93
where these authors compare the model and empirical distributions of $N_\mathrm{H}$ and
$k$. In fact, the dust-to-gas ratios $k$ is defined by FP93 as an extinction per unit
\ion{H}{i} column density and is therefore almost equivalent to a metallicity, as one
can see in  our Eq. (\ref{AlamHzeta}). 
The fact that
the metallicity distribution is much better constrained
by the observations than the  $k$ distribution   is an advantage of our approach.  
In FP93 the model distribution of $N_\mathrm{H}$  is derived 
using an exponential \ion{H}{i} radial profile;
in our work the model
is   meant to be the simplest analytical approximation
of the true distribution in the range probed by the
observations. 

\subsection{The \ion{H}{i} column density distribution}
 
From the analysis of  the \ion{H}{i} column density distribution
we derive two main results.  
First, the typical value of  $\beta$ that we find, $\beta \simeq 1.5$, 
is remarkably
similar to that measured in quasar absorbers of lower column 
densities, not affected by obscuration bias
(Tytler 1987; Petitjean et al. 1993; Storrie-Lombardi \& Wolfe 2000).
%
Second, the   biased distribution
$f^\mathrm{b}_{N_\mathrm{H}}$ 
derived from a simple power law
successfully reproduces  
the shape of the empirical distribution. 
%
Both results indicate  that the fast decline of the number of absorbers
observed in the DLA regime is due to the obscuration effect,
the true distribution of column density being instead consistent
with a simple power law, as for the majority of quasar absorbers. 
Quite interestingly, the typical
values of $\beta$ that we derive are intermediate between the predictions of
the exponential profile model of FP93, $\beta \sim 1$, and the values
expected from the mass-spectrum of interstellar clouds, $\beta \ga 2.5$
(see Section 5.2.2).  This seems to indicate that both effects should
be taken into account in the full theoretical treatment of the true distribution
$f_{N_\mathrm{H}}$.  
  The hydrodynamic simulations of DLAs by Cen et al. (2003)
yield a slope slightly below 2, broadly consistent with our results.  

The approximation of the true distribution with a simple power law   
 should not be extrapolated
 in the range of 
 column densities not sampled by the observations
 ($\log N_\ion{H}{i} \gsim 22$). 
An intrinsic faster decline in that range must occur in order to avoid
an infinite value of $\Omega_\mathrm{DLA}$ (see Section 6.5). 
A faster decline is expected  in the exponential profile model  
of disk gas distribution when  $N_\ion{H}{i}$
approaches the central value $N_{\bot 0}$ (FP93).
Also the onset of physical mechanisms 
specific of very high density environments 
may induce a sharp drop of the distribution  at
 very values of  $N_\ion{H}{i}$ (Schaye 2001). 
The present results suggest that the genuine fast decline    
of the $N_\ion{H}{i}$ distribution may lie beyond the range
probed by the observations.  
In principle, one could model this intrinsic decline with a Schechter function,
but the  position of the "knee" would be unconstrained.  
In the following we introduce        
an upper cutoff in the column density distribution  
when we need to estimate quantities affected by the
high end of the $N_\ion{H}{i}$  distribution. 
 
%

\subsection{The metallicity distribution}
  
%
The results on the metallicity distribution depend on the adopted model
for the shape of the distribution. As discussed in Section 5.3, we
believe that the results obtained from a Schechter function
are more conservative and reliable than those derived from other functions
(polynomials, lognormal or other functions parametrized in terms
of "centroid" and "width"). 
In addition to the considerations given in Section 5.3 we note here that
the position of the   "knee"  $Z_\star$ is sufficiently well  constrained
by the observational data (Fig. \ref{fits_mw}{\bf b}).
%

\begin{figure}
  \centering
 \includegraphics[width=7.cm,angle=0]{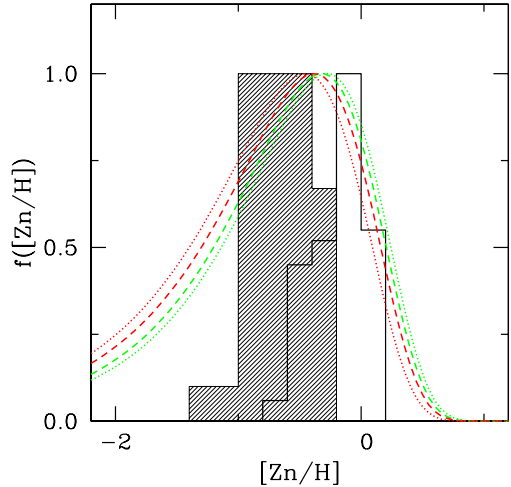}
\caption{ 
Distribution functions of the logarithmic metallicity
normalized to unity. Curves: 
true distribution of DLAs at $z \simeq 2.3$ inferred
with our method. 
Dashed (dotted) lines:  MW-type (SMC-type) extinction;
red (green): results for the $r^\prime$ ($g^\prime$) band.
Histograms: Milky-Way
stars belonging to the thick disk (shaded area)
and thin disk (thick line)
populations. 
}
 \label{metdis}
\end{figure}

The main result of the analysis of the metallicity distribution
is that a non-negligible fraction of DLAs   exist
with near solar metallicity.  
%
 %
In previous work
the metallicity distribution of DLAs has been
compared with that  of different stellar populations of the Milky Way  
to cast light on the nature of DLA galaxies. 
Pettini et al. (1997) found that 
the  [Zn/H] distribution observed in DLAs does not
resemble that of any Milky-Way population.
%
The true metallicity distribution of DLAs inferred in the present work
changes significantly  these conclusions.
In Fig. \ref{metdis} we compare the DLA distributions obtained
for the best-fit solutions listed in Table 1 with the  stellar
distributions published by Wyse \& Gilmore (1995)
for the thick and thin disk of the Milky Way
(the same template used by Pettini et al. 1997). 
One can see that the true DLA distribution  
envelops   those   of both disk populations. 
This result is consistent with the paradigm that DLA galaxies
are progenitors of present-day disk galaxies, even though
there is plenty of room in the distribution for contributions
from metal-poor galaxies. 
We cannot derive more information from these results 
because the exact shape of the true distribution is  
uncertain, particularly in the tails, owing to the
poor statistics of the surveys. 
In any case, the present results bring fresh support to
the most commonly accepted paradigm on the
nature of DLA galaxies, suggesting that most disks
may already be
in place at $z \simeq 2.3$, 
consistent with what happened in the Milky Way.

\subsection{The total obscuration fraction}

In Table 1 we list the total obscuration $\Phi_{22}$ 
computed adopting $N_\ion{H}{i}=10^{22}$ atoms cm$^{-2}$
as upper limit of integration\footnote{
The highest measured column density in a DLA system
is currently $N_\ion{H}{i}=10^{21.85}$ atoms cm$^{-2}$  (Prochaska et al. 2003).
Column densities up to $10^{22}$ atoms cm$^{-2}$ are found 
in the Milky Way ISM (Dickey \& Lockman 1990). 
}
in  Eq. (\ref{BigPhi}). This is equivalent to truncate
the \ion{H}{i} distribution  in the range of column densities not probed
by the observations. 
The true obscuration factor might be higher than
$\Phi_{22}$ if DLA systems exist also beyond such limit. 
An estimate of this uncertainty can be appreciated in Table 1, where
we also give $\Phi_{23}$, obtained integrating the  
  distributions up to  $N_\ion{H}{i}=10^{23}$ atoms cm$^{-2}$.

The total obscuration fraction   lies in the range
$\Phi_{22} \sim 0.33/0.52$
for the magnitude limit typical of high-resolution surveys
($m_\ell \sim 19.0$).
 For spectroscopic surveys of moderate resolution,
with $m_\ell \sim 19.5$, we obtain $\Phi_{22} \sim 0.27/0.46$. 
These figures are  underestimated by only $\simeq 3/5\%$ if DLAs
exist with the same power law,
up to $N_\ion{H}{i}=10^{23}$ atoms cm$^{-2}$.

The fraction of missing systems that we find is consistent
with the range 0.23 to 0.38 estimated 
by  PF95 at $z \sim 3$. 

Our estimate of total obscuration fraction  can be
compared with  that obtained by Ellison et al. (2001)
from the analysis of an unbiased sample of radio-selected quasars.
By comparing their number density of DLAs, $n(z)$, with that obtained by
Storrie-Lombardi \& Wolfe (2000) from    
  an optically-selected (biased) sample, Ellison et al.
conclude that the unbiased $n(z)$  is about $\sim 50\%$ larger
than the biased one. 
Since the Storrie-Lombardi \& Wolfe survey has a magnitude
limit typical of moderate resolution surveys, this result should be
compared with our obscuration fractions derived for $m_\ell = 19.5$,
which indicate that the true number is between $\sim 40\%$ and $\sim 90\%$
larger than the biased one. The general agreement with the result
found by Ellison et al. represents an important test of validity
of our procedure.

\begin{figure*}
  \centering
 \includegraphics[width=7.cm,angle=0]{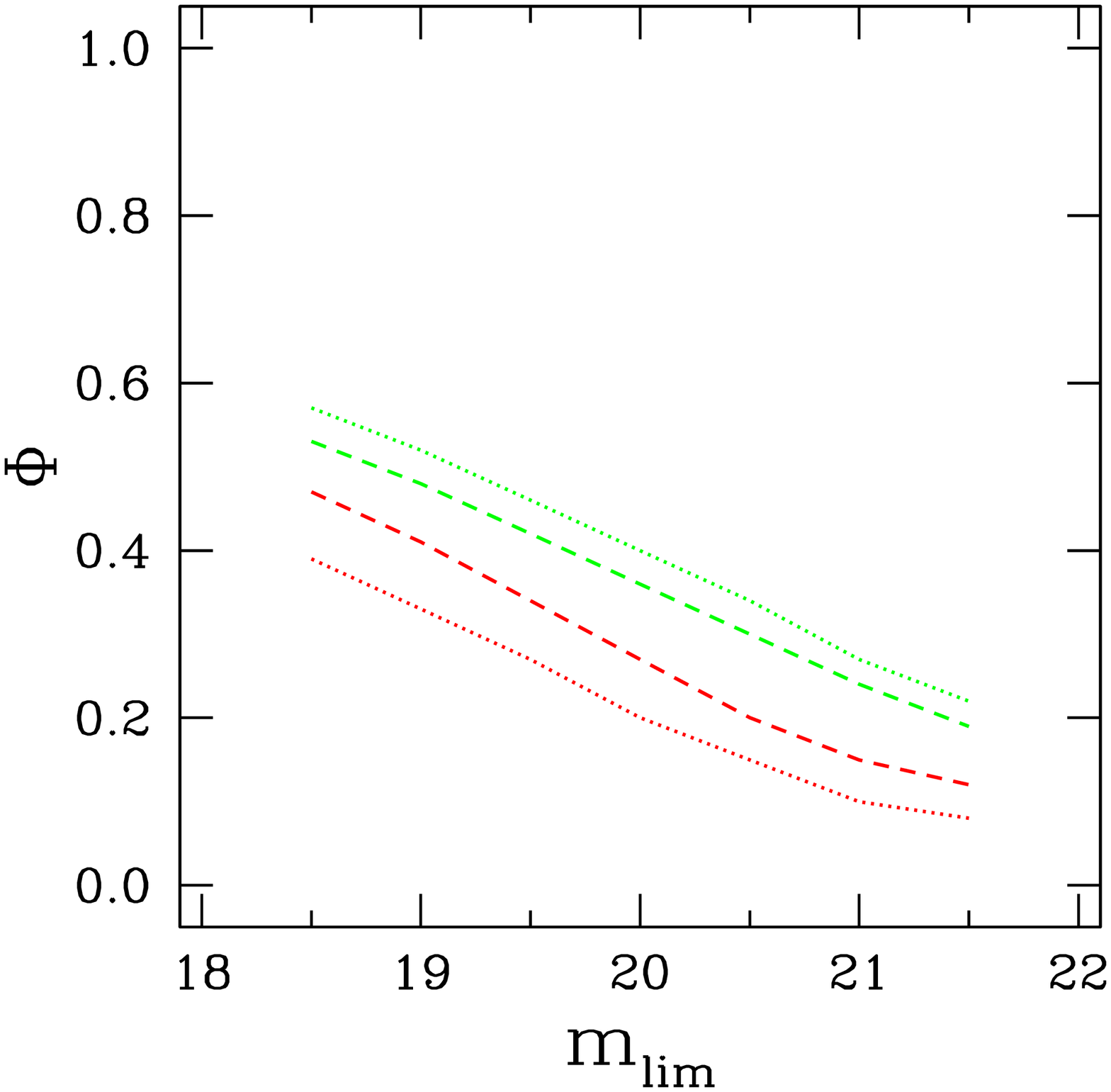}
  \hspace{0.2cm}
 \includegraphics[width=7.cm,angle=0]{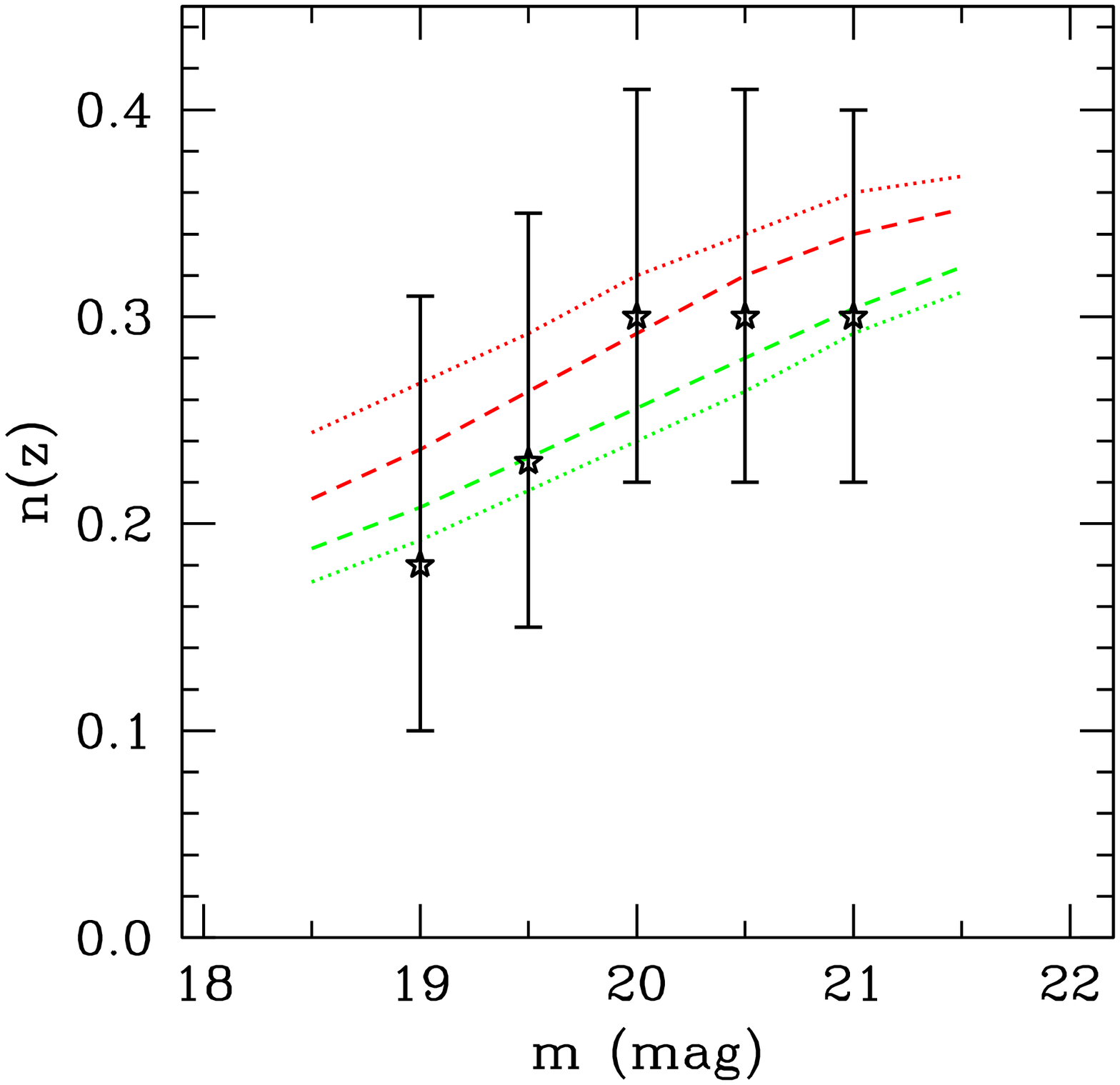}
\caption{ {\bf (a)} Fraction of obscured DLAs  and
{\bf (b)} number density of DLAs predicted as a function of the
limiting magnitude for the best-fit solutions of Table 1. 
Dashed and dotted lines:  MW-type and SMC-type extinction, respectively.
 Red and green: results for the $r^\prime$ and $g^\prime$ bands.
 Stars with error bars: CORALS results from Ellison et al. (2001). 
%
 }
 \label{maglimPhi}
\end{figure*}

\subsection{Number density versus limiting magnitude}

One prediction of the obscuration bias is that the
fraction of missed DLAs must decrease with increasing limiting magnitude
of the survey. 
The variation predicted by our computations  
is shown in Fig. \ref{maglimPhi}a for the different
cases considered in Table 1.  
The   obscured fraction decreases, but does not vanish
at least up to $m_\ell = 21.5$, where $\Phi_{22} \sim 0.1/0.2$. 
This variation of   $\Phi$ versus $m_\ell$ offers an important
observational test 
based on the measurement of the number density
 of DLAs, $n(z)$, 
at increasing values of  $m_\ell$   in an unbiased sample. 
Ellison et al. did find an increase of $n(z)$ with $m_\ell$
which supports the existence of the bias. 
The number density increases in the range   $19 \la m_\ell \la 20$,
but flattens at $m_\ell \ga 20$, while we predict
that it should keep increasing at least up to   $m_\ell \la 21.5$. 
However, taking into account the experimental error bars
  there is no real  discrepancy.
To demonstrate this, we plot in Fig. \ref{maglimPhi}b 
   $n(z) = n^\mathrm{t}(z) \, (1-\Phi)$
for an assumed total density $n^\mathrm{t}(z)=0.4$. 
One can see that these predictions agree with the measurements
of $n(z)$ performed by Ellison et al. 
If  $n^\mathrm{t}(z)=0.4$, the true number of
systems   would be $\sim 90\%$ larger  than the   biased $n(z)$ 
measured by Storrie-Lombardi \& Wolfe.  
This figure of obscuration is still consistent with our estimate.
The only "disagreement" with Ellison et al. would be on the interpretation
of the $n(z)$ versus $m_\ell$ plot, for which we claim
that there is a steady increase rather than a plateau.  
More stringent measurements are crucial for clarifyng
this issue.  
It is clear, in any case, that the combination of our treatment
of the bias with studies of unbiased samples offers 
a powerful tool for a quantitative estimate of the obscuration effect.

\subsection{ The gas content of DLAs }

By using our  best-fit $N( \ion{H}{i})$   distribution functions  
we can   compute the total contribution of the gas in DLAs
to the critical density of the Universe, 
\begin{equation}
\Omega_\mathrm{DLA} =
{ H_0 \mu \, m_\mathrm{H} \over c \, \rho_\mathrm{crit}} \,
\int^\infty_{N_\mathrm{min}} 
N_\mathrm{H} \, f_{N_\mathrm{H}} \, d N_\mathrm{H} ~.
\label{Omega}
\end{equation}
If $f_{N_\mathrm{H}} $ is a power law with $\beta < 2$, as in our case, 
the integral diverges and we must introduce an upper cutoff.
In Table 1 we give the values $\Omega_{22}$ and $\Omega_{23}$
obtained integrating up to $N_\mathrm{H} = 10^{22}$ and $10^{23}$
atoms cm$^{-2}$, respectively. The value $\Omega_{22}$ 
is conservative since the true distribution $f_{N_\mathrm{H}}$ may
extend above  $N_\mathrm{H} \sim 10^{22}$ cm$^{-2}$.
The values that we find for   different    input parameters
lie in a relatively narrow range, namely 
$1.9 \times 10^{-3} \leq  \Omega_{22} \leq 3.1 \times 10^{-3}$.

To estimate the missed fraction of \ion{H}{i} mass, we compare these values
of $\Omega_{22}$ with  
$ \Omega^\mathrm{obs}_\mathrm{DLA} \simeq 0.9 \times 10^{-3}$,
the  biased  value   obtained  integrating our   \ion{H}{i} sample 
with the commonly adopted Schechter function   (\ref{schechter}).   
Since   $\Omega_{22}$ is a conservative estimate,
this comparison indicates that at $z \sim 2.3$
the true comoving mass of DLAs is at least
a factor of 2 higher than that obtained by magnitude-limited
surveys.  

The values of $\Omega_{22}$ that we derive agree well  
with the value $ \Omega_\mathrm{E01} \simeq 2.6 \times 10^{-3}$
measured by Ellison et al. (2001) from the analysis of their "dust-free" sample. 
 In addition, they agree with the value 
$\Omega_\mathrm{DLA} \simeq 3 \times 10^{-3}$
derived at $z \simeq 2.3$ by Cen et al. (2003)
 from their hydrodynamic simulations.

%


%


%

\subsection{ The metal  content of DLAs }

The column-density weighted metallicity of DLA systems 
is used to measure the degree of metal 
enrichment of the population of DLAs as a whole (Pettini et al. 1999) and  
to estimate the mean cosmic metallicity of the high-redshift
Universe (Pei \& Fall 1995; Cen et al. 2003).    
The expression commonly used in literature to measure the weighted metallicity
from a finite set of column densities, 
 $ <\! (\mathrm{Zn/H})_\mathrm{DLA} \!> =
{  ( \sum_i N_{\mathrm{Zn},i}  ) /  (\sum_i N_{\mathrm{H},i} )}$,
can be put in the form
\begin{equation}
<\!  ( \mathrm{Zn} / \mathrm{H}  )_\mathrm{DLA} \!> =
{  
\int N_\mathrm{Zn} ~ f_{ N_\mathrm{Zn} } ~ d N_\mathrm{Zn}  
 \over
\int N_\mathrm{H} ~ f_{ N_\mathrm{H} } ~ d N_\mathrm{H}  
}~,
\label{MeanZnH}
\end{equation}
to measure the same quantity from the frequency distribution functions
$f_{N_\mathrm{Zn} }$ and $f_{N_\mathrm{H} }$ normalized to unit area. 
If the metallicity $Z$ and the \ion{H}{i} column density are independent
variables, as we assume in our mathematical formulation, 
then  it is easy to show that
$<\! (\mathrm{Zn/H})_\mathrm{DLA} \!>$ equals the expectation value 
$<\! Z \!> =  \int Z ~ f_Z ~ d\, Z $ estimated from the   
distribution $f_Z$ normalized to unit area.

In Table 1 we list the     mean metallicity, normalized in
the usual form $\log (<\! Z \!>/ Z_{\sun})$,  
obtained for a Schechter metallicity distribution.
The mean metallicity
lies between $-0.44$ and $-0.27$ dex below
the solar level for the different dust models and photometric bands
considered in Table 1. 
Considering the total error of $\sim -0.3$ dex in our estimate
(fit errors plus statistical errors of the empirical distributions)
we conclude that the mean metallicity is    
  $ \gsim -0.7$ dex at 1 $\sigma$ level.
These values are   higher
 than the direct measurements of weighted metallicity of DLAs,
 which give  $\sim -1.1$ dex 
at the typical redshift of our sample (Pettini et al. 1999; Prochaska et al. 2003). 
This large difference is probably be due to the fact 
that the weighted metallicity is extremely
sensitive to  the number of  high column-density
DLAs  included in the average, which are  
the DLAs  more affected by   obscuration.   
%
Our results are marginally consistent   with the mean
weighted metallicity  
  $\left[ <\!  \mathrm{(Zn/H)_{DLA}} \!> \right] = -0.88 \pm 0.21 $ dex obtained from the  
  CORALS metallicity survey (Akerman et al. 2005).

    \begin{figure*}
   \centering 
     \includegraphics[width=5.7cm,angle=0]{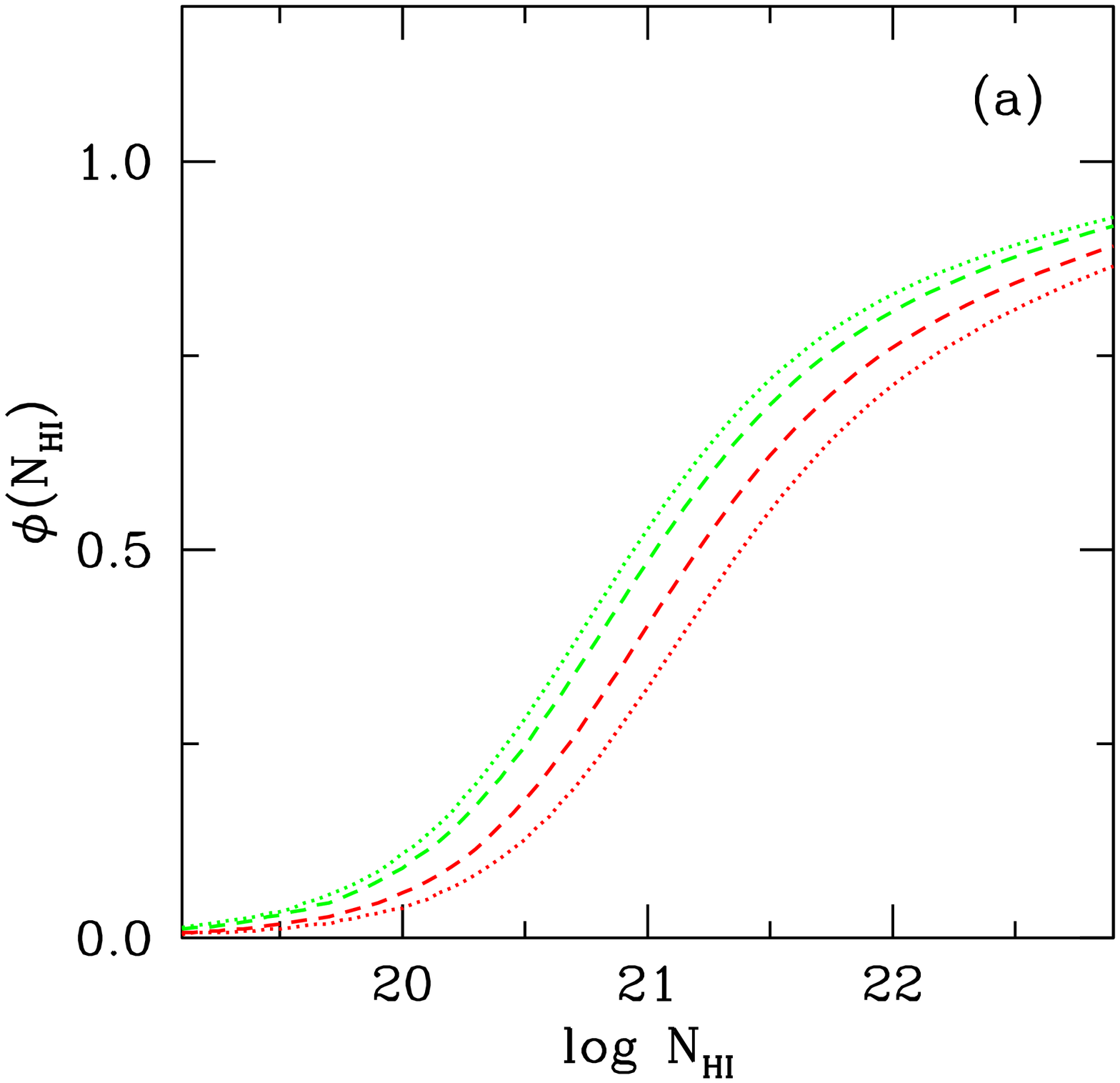} 
 \hspace{0.2cm}
  \includegraphics[width=5.7cm,angle=0]{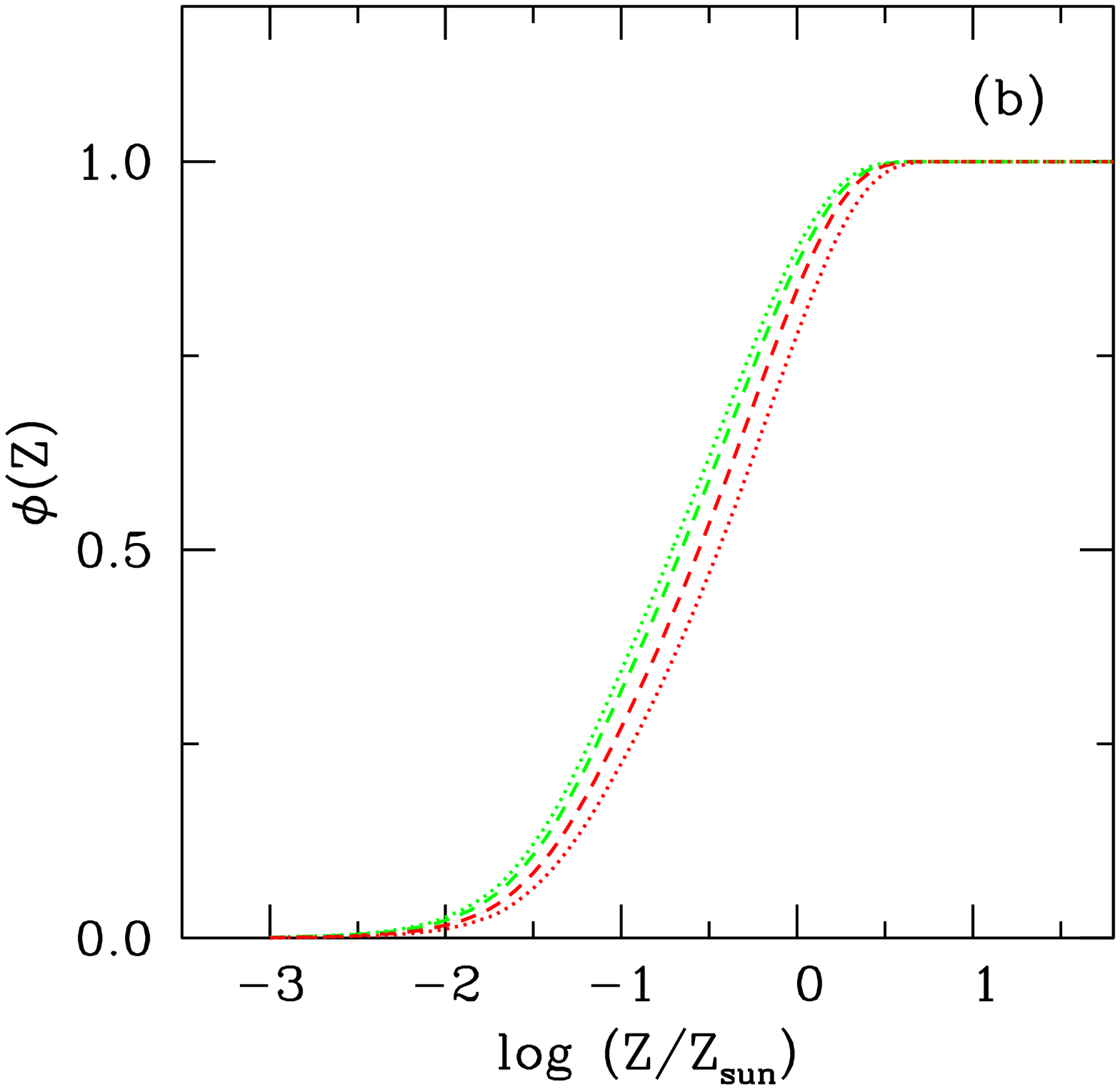} 
 \hspace{0.2cm}
  \includegraphics[width=5.7cm,angle=0]{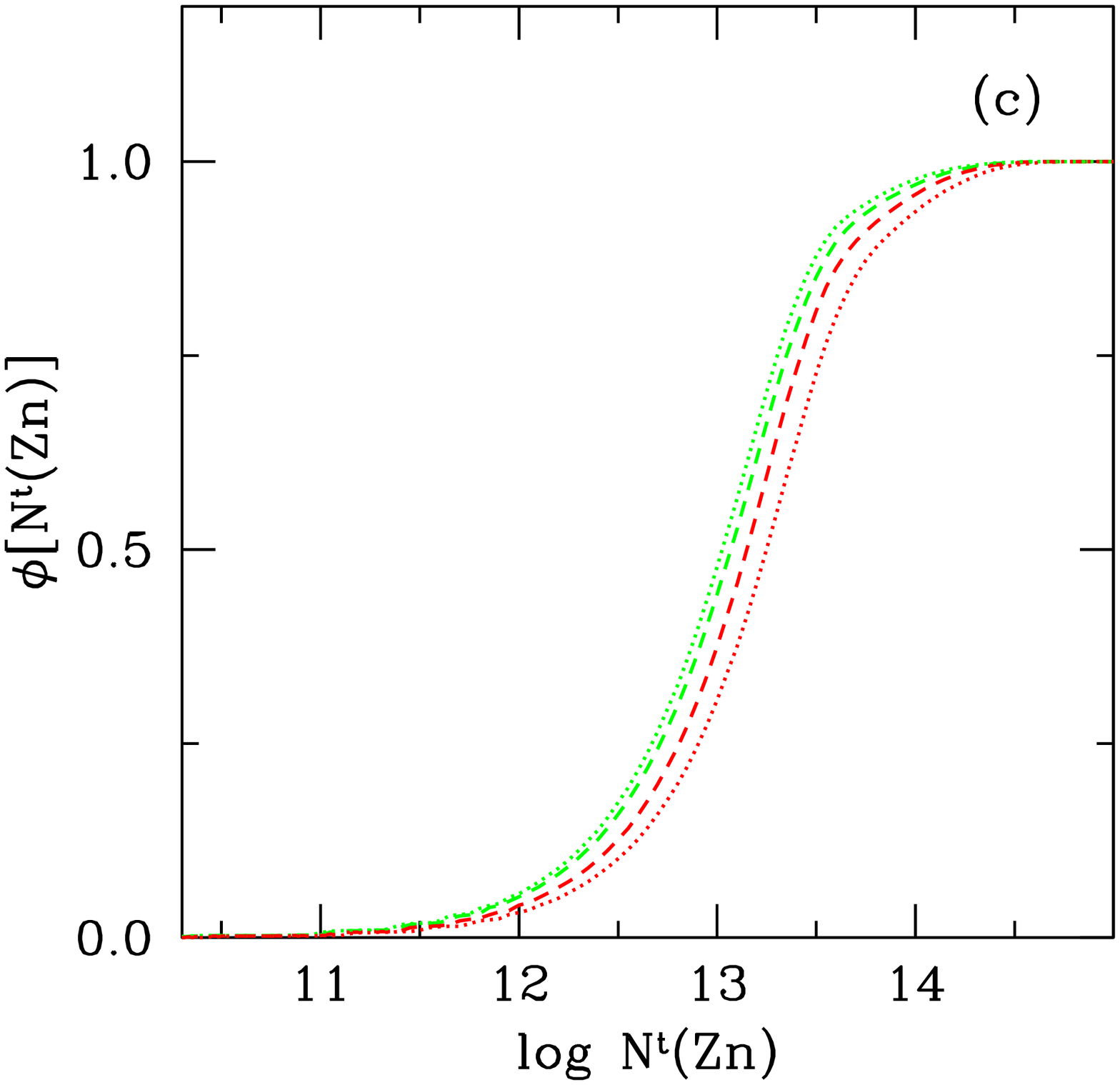} 
  \caption{  Fraction of obscured DLAs as a function
  of  {\bf (a)} \ion{H}{i} column density,
  {\bf (b)} metallicity $Z$, and {\bf (c)} \ion{Zn}{ii} column density
  for the best-fit solutions of Table 1.
 Dashed and dotted lines:  caption as in Fig. \ref{maglimPhi}.
  }
 \label{plotPhi}%
    \end{figure*}

Quite interestingly, the high mean metallicity that we derive
may help to solve the "missing metal" problem 
pointed out by  Wolfe et al. (2003) 
in their work on the Star Formation Rates (SFRs) in DLAs: 
the mass of metals produced by $z \simeq 2.5$, inferred from their SFRs,
is 30 times larger than detected  in absorption in DLAs. 
%

\subsection{Observational thresholds due to obscuration bias }

Our method allows us to estimate, for the first time,  
the  fraction of obscured  DLAs as a function
of \ion{H}{i} column density and metallicity $Z$, and also as
a function of the total zinc
column density. 
These  obscuration fractions
$\phi_\ell(N_\mathrm{H})$, $\phi_\ell(Z)$
and $\phi_\ell(N^\mathrm{t}_\mathrm{Zn})$, shown in Fig. \ref{plotPhi},
should only be interpreted in a statistical sense. However, they
provides us a powerful indication of how the statistical
distributions of DLAs are distorted by the obscuration bias.
In particular, they provide a means of quantifying the existence of
observational cutoffs induced by the bias itself. 
 
The analysis of $\phi_\ell(N^\mathrm{t}_\mathrm{Zn})$, shown in panel {\bf (c)},
gives a quantitative explanation for the lack of DLAs systems above the
obscuration threshold in \ion{Zn}{ii} column density
originally proposed by Boiss\'e et al. (1998). We estimate 
that    over $60\%$ of DLAs are missed
at $N^\mathrm{t}_\mathrm{Zn} = 10^{13.2}$ atoms cm$^{-2}$,
and this fraction rises rapidly to   $\sim 95\%$ if the 
$N^\mathrm{t}_\mathrm{Zn}$ increases by $\simeq 0.5$ dex.  
The threshold is rather independent of the adopted input parameters.
Only for  peculiar
extinction curves without UV rise (gray extinction) we may expect that
the threshold can be crossed. 

We can use our results to estimate how many DLAs should be observed
in an unbiased survey in order to  detect cases with high metal column density. 
From the typical distribution $f_{N_\mathrm{Zn}}$ that we infer,
we estimate that the number of DLAs      
  in the range between the detection limit
and the Boiss\'e's threshold ($11.5 \lsim \log N(\ion{Zn}{ii}) \lsim 13.1$) 
is about 24 times larger than the number of systems
in a range of similar extension above the threshold ($13.1 < \log N(\ion{Zn}{ii}) \lsim 14.7$). 
This means that one is not guaranteed to detect one DLA above the threshold 
even with $\sim 30$ unbiased spectroscopic observations. 
This is currently an observational challenge because the high extinction of the DLAs above
the threshold makes very hard to perform high-resolution spectroscopy, even
for the quasars which lie at the bright end of the true distribution
of apparent magnitudes. 

A natural consequence of the cutoff at high metal column densities,
in conjunction with the artificial DLA threshold 
$N_\ion{H}{i} \geq 10^{20.3}$ atoms cm$^{-2}$,
is the existence of an observational cutoff at high metallicities. 
From panel {\bf (b)} one can see that about $90\%$ of systems are
missed at solar metallicity and it would be practically impossible
to detect systems with oversolar metallicity,  
if they exist. Also worth of mention is
the fast rise of the obscuration with increasing metallicity, which
distorts dramatically our perception of the metallicity 
distribution of DLA systems. The fact the different curves plotted
in panel {\bf (b)} lie close to each other indicates that
these results are rather
independent of the parameters adopted. 


%
Also the distribution of \ion{H}{i} column densities is
severely distorted by the extinction bias, as shown in
panel {\bf (a)}. The curves of $\phi(N_\mathrm{H})$ obtained
for different parameters are not tightly close to each other, but
the general behaviour is quite similar in all cases. 
The obscuration  starts from a low value 
in the sub-DLA   regime and rises steadily  
over the DLA regime,  
approaching  $\sim 80\%$  at $N_\ion{H}{i} \sim 10^{22}$ atoms cm$^{-2}$.
This is the approximate value of the observational threshold
for detection of high column density DLAs. 
The rise of    $\phi(N_\mathrm{H})$,
in conjunction with the genuine decrease of the true   distribution
$f_{N_\mathrm{H}}$,     provides a natural explanation
for the  non-detection of DLAs 
at $N_\ion{H}{i} \ga 10^{22}$ atoms cm$^{-2}$.  
This  makes it unnecessary to invoke the sudden transition
from atomic to molecular hydrogen proposed by Schaye (2001)
in order to explain  this empirical cutoff.   
In any case, if  fully molecular clouds exist as 
predicted by Schaye, they  would   be
characterized by a high column density of dust grains.
The fact that  quasar absorbers with $N(\mathrm{H_2}) > N(\ion{H}{i})$
have not yet been detected may indicate that also this type of absorber
is missed due to    obscuration.

\section{Summary and conclusions}

 Starting from
  the theoretical relation between the  interstellar extinction, $A_\lambda$,
and  the  column density of a volatile metal, $N_{\mathrm{X}_v}$,
we investigated the detailed 
relation between $A_\lambda$ and  $N_\mathrm{Zn}$ 
  in DLAs. 
%
%
We derived in this way  the Eq. (\ref{AlamHzeta}), which gives
the quasar extinction in the observer's frame  as a function of 
the DLA hydrogen column density, $N_\mathrm{H}$, metallicity,
$Z = N_\mathrm{Zn}/N_\mathrm{H}$, fraction in dust of iron, $f_\mathrm{Fe}$, and a factor $G$, 
which describes   the dust grain properties. 
 For the first time, the metallicity evolution of the 
dust-to-metal ratio in DLAs is explicitly accounted for in this type of relation,
using an expression  $f_\mathrm{Fe} = f_\mathrm{Fe}(Z)$ 
obtained from a previous study of depletions (Vladilo 2004). 
We argue that the factor $G$ may not   vary
substantially in   interstellar environments 
with moderately low metallicity ($Z \ga 0.1 \, Z_{\sun}$)
and derive
a value of 60\% the Milky-Way value   for SMC-type dust. 
Possible variations of $G$ at the early stages of chemical evolution
do not affect the extinction   
because $f_\mathrm{Fe}(Z)$ vanishes
when $Z \ll Z_{\sun}$.

We searched for empirical evidence of the rise
of the   extinction with  metal column density in DLAs. 
We did find an increase of the quasar magnitude with $N_\mathrm{Zn}$, 
consistent with this expectation
(Fig. \ref{Vmaghist}). 
 
Starting from Eq. (\ref{AlamHzeta}), 
we derived a mathematical formulation aimed at  estimating 
how the frequency distributions
of \ion{H}{i} column densities and metallicities of a sample of DLAs
are biased as a result of the obscuration 
generated by the DLAs of the same sample.
We ignore 
the   obscuration due to low-redshift DLAs 
along the same line of sight, showing that this contribution
 is relatively low. 
At variance with the formulation of PF95, where
a constant dust-to-metal ratio was adopted for all DLAs,   we adopted
a
metal-dependent dust-to-metal ratio $f_\mathrm{Fe}(Z)$. 
We cross-checked the validity of our formulation making use
of the equations derived by FP93 for the case of a power law
quasar luminosity function and constant dust-to-gas ratio (Section A.1.2).

%
We presented a practical procedure for 
recovering the unbiased distributions of column densities and
metallicities, $f_{N_\mathrm{H}}$ and $f_Z$, 
using the empirical distributions obtained from magnitude-limited surveys.
%
The unbiased distributions are modelled with simple analytical
expressions in parametric form.    
%
The bias induced by   the extinction
on the quasar 
magnitude distribution is accounted for self-consistently by the procedure. 
A unique characteristic of the method is
the possibility of recovering the true metallicity distribution
starting from an educated guess of the functional form of the distribution.
We have shown that using a Schechter function 
for the metallicity distribution appears to provide  more conservative
and reliable results
than other functions (e.g. lognormal or polynomials) for the estimate
of the bias (Section 5.3). 
%


We applied  our method to the   sample of DLAs 
in the redshift interval $1.8 \leq z \leq 3$, 
with mean redshift  $<\!\! z \!\!> \simeq 2.3$,
where the bulk of   spectra are available. 
%
%
%
%
The  extinctions were computed in the $g^\prime$ and $r^\prime$
photometric bands of the SDSS, considering both a 
MW-type and an SMC-type 
average extinction curve.
%
We found that the effect of the obscuration
on the quasar magnitude distribution is modest (e.g. Fig. 5).
On the other hand,
  the bias plays an important role
in shaping the statistical distributions of DLAs. 
%

The unbiased distribution function of \ion{H}{i} column densities is  
successfully approximated by a power law 
$f_{N_\mathrm{H}} \propto N_\mathrm{H}^{-\beta}$,
with $\beta \simeq 1.5$. 
This simple law, in conjunction with the bias effect, 
is able to fit the observed decline of the   number DLAs
with $N_\mathrm{H}$
   up to $N( \ion{H}{i}) \sim 10^{21.8} $ atoms cm$^{-2}$. 
The faster drop  of the distribution  expected at very high column
densities (FP93; Schaye 2001) probably lies outside the
range currently probed by the observations. 
%
The   slope   $\beta \simeq 1.5$ is remarkably similar to the value 
which characterizes most of quasars absorbers with lower column densities,
for which the extinction bias is negligible. 
%
%
The value of the slope  suggests that the mass spectrum
of interstellar clouds may play an important role in shaping
the distribution.

The observed distribution of metallicities is  
reproduced modelling the unbiased distribution with a 
Schechter function  
$f_Z ~ \propto ~ (Z/Z_\ast)^\alpha ~
 e^{-Z/Z_\ast} $, with $\alpha \simeq -0.4$ 
 and $\log ( Z_\ast/Z_{\sun} ) \simeq -0.15$ dex. 
Once converted into logarithmic scale, this function  
peaks at [Zn/H] $\sim -0.3/-0.4$ dex, indicating that
a non-negligible fraction of DLAs exist with near solar metallicity.
%
%
%
The true metallicity distribution of DLAs envelops the distributions
of the thick-disk and thin-disk Milky-Way  stellar populations.  
This result is in line with the paradigm that DLA galaxies
are progenitors of present-day disk galaxies. However, in
the inferred  distribution there is also room for
a relevant contribution from metal-poor galaxies.  

The  mean weighted metallicity that we derive ,
 $ \simeq -0.4/ \! -0.3$ dex,    
is significantly higher than 
the   column-density weighted metallicity
$<\!\! \mathrm{[Zn/H]} \!\!> = -1.1$ dex
measured at $z \simeq 2.3$. 
 This high value of mean metallicity may  help
to solve the  discrepancy between the metallicity observed in DLAs
and that predicted on the basis of their SFRs (Wolfe et al. 2003). 

The fraction by number of obscured DLAs, $\Phi$, decreases
with increasing limiting magnitude of the survey, $m_\ell$.
%
We obtain    $\Phi \ga 0.4/0.6$ at $m_\ell \la 18.5$, a limit
representative of  high resolution spectroscopic surveys, 
carried out with 4-m class telescopes. 
The corresponding figures
for high-resolution surveys in
10-m class telescopes  are  $\Phi \simeq 0.35/0.55$
($m_\ell \simeq 19.0$).

%
%

An important result of our work is the explanation (Fig. \ref{plotPhi}c)  of the
observational limit $N_\mathrm{Zn} \sim 10^{13.2}$ atoms cm$^{-2}$ 
(Boiss\' e et al. 1998) in terms of very simple physics (Section 2), with no tuning
of the local dust parameters. 
Alternative explanations  would require 
  an additional mechanism   able
to produce the exact same limit predicted by the extinction. 
The simplicity of the extinction model  
favours the existence of the obscuration. 
The existence of the threshold $N_\mathrm{Zn} \sim 10^{13.2}$ atoms cm$^{-2}$ 
is fundamental for reconciling 
the predictions of galactic models 
(Prantzos \& Boissier 2000, Hou et al. 2001; Churches et al. 2004)
and cosmological simulations (Cen et al. 2003, Nagamine et al. 2004)
with the observations of DLAs. 
Only for gray extinction curves, without UV rise,
we may expect that the threshold 
can be crossed. 
Our results on the magnitude of the obscuration effect 
are broadly consistent with those 
presented by  FP93 and PF95, 
with some important differences. 
The modest bias of quasar statistiscs that we find is at variance with
the predictions of FP93, which allowed up to 70\% of quasars
to be obscured. 
Contrary to FP93,  we do not expect a significant contribution
of DLA extinction at redshift $z \gsim 3$
in spite of the   rise of $A_\lambda$ with $z$. 
In fact, our model of dust fraction  $f_\mathrm{Fe}(Z)$ vanishes
at very low metallicity so that the extinction  (\ref{AlamHzeta})  
vanishes at $z \gsim 3$, when the metallicity $Z \lsim 0.01 Z_{\sun}$.


A crucial test of  our results is the comparison with 
studies of radio-selected samples of quasars/DLAs.  
Our determinations are consistent with the estimates
of   obscuration fraction, number density   and $\Omega_\mathrm{DLA}$ 
obtained from the    CORALS
survey (Ellison et al. 2001).
Our prediction of mean metallicity is marginally consistent
with the recent estimate of mean weighted metallicity
of the CORALS sample
(Akerman et al. 2005). 

%
%

Accurate estimates of the obscuration effect will be possible
as soon as the  samples with metallicity
measurements   become sufficiently large for   
precise statistical analysis. 

 \begin{acknowledgement}
This work has benefitted from interactions with  
 Irina Agafonova, Patrick Boiss\'e, Miriam Centuri\'on, Sergei Levshakov, Paolo Molaro, Pierluigi Monaco,
 Sara Ellison and Paolo Tozzi. 
 We warmly thank  
 Arthur Wolfe for his remarks on the first part of the manuscript.
\end{acknowledgement}

 
\appendix

\section{Mathematical formulation}

\subsection{ True and biased distributions }

%
To derive our mathematical formulation
we define the frequency distributions of DLAs and quasars 
most appropriate for taking into account the causes and effects
of the extinction bias. 
To account for the DLA extinction via Eq. (\ref{AlamHzeta})
we are interested in the number   of DLAs 
per unit hydrogen column density and metallicity.
To account for  the dimming of the quasar  we are interested
in the number of quasars per unit of apparent magnitude.
The apparent magnitude, $m$, can be directly compared with the limiting magnitude of the observational survey, $ m_{\ell}$. 


We call $h(N_\mathrm{H},Z,z) \, d N_\mathrm{H} \, dZ\, dz \, d\omega$
the   number  of neutral hydrogen layers
  with column density
$N_\mathrm{H} \in (N_\mathrm{H}, N_\mathrm{H}+dN_\mathrm{H})$ and
metallicity $Z \in (Z, Z+d \,Z)$ located in the  
spherical shell of redshift $z \in (z,z+dz)$
in the infinitesimal solid angle $d\omega$ 
in a random direction
(we assume that the gas layers are distributed
isotropically). 
 
We call $q(m,  z_e) \, dm  \, dz_e $
the true number  of quasars  
with apparent magnitude  $m \in (m,m+dm)$  
in the   shell of
emission redshift $z_e \in (z_e, z_e+d z_e)$ all over the sky 
(we assume an isotropic distribution of quasars). 
We  define        
$n(m; z, z_{e}^\mathrm{max})
\equiv \int_{z}^{z_{e}^\mathrm{max}} q (m,  z_e) \, d z_e$,
where $z$ is the absorption redshift of a foreground \ion{H}{i} layer    
and $z_{e}^\mathrm{max}$ the maximum redshift of the quasars in the survey. 
For simplicity of notation we omit hereafter
the dependence on $z_{e}^\mathrm{max}$. 
Therefore  $n(m; z) \, dm$ is the
total number  of quasars  with magnitude  $m \in (m,m+dm)$
observable beyond   redshift   $z$  all over the sky.
This is the true number
in absence of quasar obscuration.

We call $\delta\omega_\mathrm{q} (z)$
the average
solid angle subtended in the observer's optical band
by quasars in the redshift interval $z < z_{e} \leq z_{e}^\mathrm{max}$.

With the above definitions, the number of DLAs 
with    
$N_\mathrm{H} \in (N_\mathrm{H}, N_\mathrm{H}+dN_\mathrm{H})$,
 $Z \in (Z, Z+d \,Z)$ and redshift $z \in (z,z+dz)$
in front of one single background quasar  with $m \leq m_{\ell}$ is 
$$h(N_\mathrm{H},Z,z) 
\, d N_\mathrm{H} \, dZ\, dz \, \delta\omega_\mathrm{q}(z) $$ 
provided $N_\mathrm{H}=N(\ion{H}{i}) \geq 10^{20.3}$ atoms cm$^{-2}$.
The number of DLAs in the same differential element    
$d N_\mathrm{H} \, dZ\, dz$
that are detectable 
in front of the $n(m; z)  \, dm$
background quasars with $m \in (m,m+dm)$ is
 $$h(N_\mathrm{H},Z,z) 
\, d N_\mathrm{H} \, dZ\, dz \, \delta\omega_\mathrm{q}(z) \times \, n(m; z)  \, dm \, ,$$ 
provided $m \leq m_{\ell}$. 
By integrating in $dm$ up to the limiting magnitude $m_\ell$
we calculate the total number of  DLAs   that are   detectable
in the survey.
We first consider the ideal case in which  no quasar extinction
  is present. 
In this case the unbiased number of detectable DLAs  
in $d N_\mathrm{H} \, dZ\, dz$ is
 \begin{eqnarray}
\lefteqn{
g_{m_\ell}(N_\mathrm{H},Z,z) \, d N_\mathrm{H} \, dZ\, dz    \equiv {} }
                  \nonumber\\
& &
{}            h(N_\mathrm{H},Z,z)   \,
\delta\omega_\mathrm{q}(z)
   \left[ \int_{\circ}^{m_{\ell}} n(m; z) \, dm \right]  d N_\mathrm{H} \, dZ\, dz    
                  \, .
\label{truedetDLAs}
\end{eqnarray}
We call $g_{m_\ell}(N_\mathrm{H},Z,z)$ the true distribution of 
detectable DLAs in $d N_\mathrm{H} \, dZ\, dz$. 

We then consider  the quasar extinction $A_{\lambda} (N_\mathrm{H},Z; z)$   
due to the DLAs  in 
$d N_\mathrm{H} \, dZ\, dz$  (Eq. \ref{AlamHzeta})
 ignoring other sources of extinction. 
From this we define the bias function 
\begin{equation}
b_{m_\ell} (N_\mathrm{H},Z, m,z) \equiv
\left\{
\begin{array}{ll}
1 & ~~ \mbox{if $m \leq{m}_{\ell} -A_{\lambda} (N_\mathrm{H},Z; z)   $ } \\
0 & ~~ \mbox{if $m > {m}_{\ell} -A_{\lambda} (N_\mathrm{H},Z; z)  $ } \\ 
\end{array} ~.
\right.
\label{BiasFun}
\end{equation} 
%
The number of DLAs  in  
$d N_\mathrm{H} \, dZ\, dz$ that,
in spite of their own extinction, are detectable 
in front of the $n(m; z)  \, dm$
background quasars    is 
$$b_{m_\ell}(N_\mathrm{H},Z, m,z) \, h(N_\mathrm{H},Z,z) 
\, d N_\mathrm{H} \, dZ\, dz \, \delta\omega_\mathrm{q}(z)  \, n(m; z)  \, dm .$$
By integrating in $dm$ we obtain the total number of   DLAs  
in   $d N_\mathrm{H} \, dZ\, dz$
that, in spite of their own extinction,  are detectable in the survey  
\begin{eqnarray}
\lefteqn{
g_{m_\ell}^\mathrm{b}(N_\mathrm{H},Z,z) \, d N_\mathrm{H} \, dZ\, dz    \equiv 
   h(N_\mathrm{H},Z,z) \,
\delta\omega_\mathrm{q}(z) \times
{} }
                  \nonumber\\
{}    &  \times
        \left[ \int_{\circ}^{m_{\ell}-A_\lambda(N_\mathrm{H},Z, z)}  \, 
        n(m; z) \, dm \right]  d N_\mathrm{H} \, dZ\, dz   \, .&   
\label{biasdetDLAs}
\end{eqnarray}

The number of DLAs in the survey that are
missed as a consequence of their own extinction is
$\left( g_{m_\ell}-g_{m_\ell}^\mathrm{b} \right) 
\, d N_\mathrm{H} \, dZ\, dz $.
The fraction of DLAs 
missed as a consequence of their own extinction is  
\begin{equation}
\varphi_{m_\ell} (N_\mathrm{H},Z, z) \,
\equiv
{ g_{m_\ell}-g_{m_\ell}^\mathrm{b} \over g_{m_\ell}}
= 1 - B_{m_\ell}  ( N_\mathrm{H}, Z,z) \, ,
\label{philost}
\end{equation}
where
\begin{equation}
B_{m_\ell}  ( N_\mathrm{H}, Z,z) \equiv
{
  \int_\circ^{m_{\ell}-A_\lambda(N_\mathrm{H},Z, z)} \,  
 n(m; z)  \, d m \,
   \over
  \int_\circ^{m_\ell}  \, n(m; z)  \,  dm 
 \label{BigBHZz}
 } \, .
\end{equation}

The ratio of the integrals of the   magnitude distribution
in Eq. (\ref{BigBHZz}) 
is reminiscent
of the ratio of the integrals of the quasar luminosity function
in Eq. (3) of   FP93.
This ratio represents, in a sense, the core relation
from which the bias effect is estimated.

The relation between the true distribution
and the distribution
biased by its own  DLAs is
\begin{equation}  
g_{m_\ell}^\mathrm{b}(N_\mathrm{H},Z,z)  
= B_{m_\ell}  ( N_\mathrm{H}, Z,z)  ~
g_{m_\ell}(N_\mathrm{H},Z,z) \, .
\label{HZzdistr}
\end{equation}
This   relation 
is independent of   the average solid angle
$\delta\omega_\mathrm{q} (z)$.   
The normalization factor of $n(m,z)$ is also
irrelevant because it is eliminated in Eq. (\ref{BigBHZz}).

To make practical use of Eq. (\ref{HZzdistr}) 
we need to understand the relation between the biased distribution
$g_{m_\ell}^\mathrm{b}(N_\mathrm{H},Z,z) $
and the observed distribution of DLAs in 
$d N_\mathrm{H} \, dZ\, dz$, that we call
$g_{m_\ell}^\mathrm{obs}(N_\mathrm{H},Z,z)$.
To derive an expression for the observed distribution
we must consider all the sources of obscuration,
including the DLAs at $z^\prime < z$ that lie in the
same lines of sight of the DLAs at redshift $z$.
If these additional sources of obscuration are negligible,
then the expression (\ref{HZzdistr}) is a good
approximation  of the observed distribution (see A.1.1), i.e.
\begin{equation}
g_{m_\ell}^\mathrm{obs}(N_\mathrm{H},Z,z) 
\simeq
g_{m_\ell}^\mathrm{b}(N_\mathrm{H},Z,z) \, .
\label{approximation}
\end{equation}
%
%
{\em 
Using the above approximation, the treatment of the obscuration bias
is self-contained, in the sense that  given the true distributions 
of column densities and metallicities at a redshift $z$, 
the observed distributions of the same quantities are uniquely determined. 
In this way we can study the  
obscuration effect in a redshift interval sampled by the observations,
even if we do not have   good statistics of the DLAs at lower redshifts. 
}
%
    
%
To compare the distributions with the observations 
we   integrate Eq. (\ref{HZzdistr}) 
in $dz$  over  the interval 
$(z_\mathrm{min},z_\mathrm{max})$ with  
$\overline{z}=(z_\mathrm{max}-z_\mathrm{min})/2$ 
close to the peak of the observed distribution.
If  the statistical functions  of interest  show a smooth variation  
inside the interval we can approximate them   
with their value at  $z=\overline{z}$.
%
We obtain         
\begin{equation}  
f^\mathrm{b}(N_\mathrm{H},Z)  
\simeq \mathcal{B}_{m_\ell} (N_\mathrm{H}, Z)  ~
f(N_\mathrm{H},Z) ~,
\label{HZdistr}
\end{equation} 
where 
 \begin{eqnarray} 
f^\mathrm{b}(N_\mathrm{H},Z) &
\equiv &  
\int_{z_\mathrm{min}}^{z_\mathrm{max}}     
g^\mathrm{b}(N_\mathrm{H},Z, z) \ dz  
  \nonumber \\ 
f(N_\mathrm{H},Z)~ &
\equiv &  
\int_{z_\mathrm{min}}^{z_\mathrm{max}}    
g(N_\mathrm{H},Z, z) \ dz 
\label{intgz}
\end{eqnarray}
and
\begin{equation} 
\mathcal{B}_{m_\ell} (N_\mathrm{H}, Z) 
\equiv 
B_{m_\ell} (N_\mathrm{H}, Z, \overline{z}) ~.
\label{BigCalB}
\end{equation}
For simplicity   we omit   the dependence of these
functions   on $\overline{z}$,
$z_\mathrm{min}$ and $z_\mathrm{max}$.

We now  transform the distribution $f(N_\mathrm{H},Z)$
into a distribution of column densities, $f_{N_\mathrm{H}}$,
and a distribution of metallicities, $f_{Z}$, defined in such a way that
$f_{N_\mathrm{H}} d N_\mathrm{H}$ is the  
true number of detectable DLAs with  column densities between
$N_\mathrm{H}$ and  $N_\mathrm{H}+dN_\mathrm{H}$ 
in the   interval $(z_\mathrm{min},z_\mathrm{max})$
and $f_{Z} \, dZ$ the true  number of detectable DLAs with
metallicity between $Z$ and $Z+d \,Z$ in the same redshift interval.  
We call $f^\mathrm{b}_{N_\mathrm{H}}$
and $f^\mathrm{b}_{Z}$ the corresponding biased distribution functions.

To proceed further we  assume that $N_\mathrm{H}$ and $Z$  are independent
variables\footnote{
A similar assumption was done by Fall \& Pei (1993; Appendix) to   
separate the distribution of column densities from that of dust-to-gas ratios.
}
 (this assumption is discussed in Section 4.4).
This implies that  
$f(N_\mathrm{H},Z) \, d N_\mathrm{H} \, d\, Z=
f_{N_\mathrm{H}} \, d N_\mathrm{H} \, f_Z \, d\, Z$. 
 We substitute this expression in the right hand of Eq. (\ref{HZdistr}). 
We then integrate that equation  in metallicity and obtain  
 \begin{equation}
\int_{0}^{\infty} f^\mathrm{b}(N_\mathrm{H},Z) \, dZ \simeq 
\left[ \int_{0}^{\infty} 
  \mathcal{B}_{m_\ell}  ( N_\mathrm{H}, Z) \,
f_{Z}  \, d\, Z \right]   ~  f_{N_\mathrm{H}} ~~.
 \label{fb_H_a}
\end{equation}
Apart from a constant factor, the left term of this equation represents
the biased distribution $ f^\mathrm{b}_{N_\mathrm{H}}$. 
The constant   
can be determined by imposing that in absence of bias
the true and biased distributions must be equal.
From this we derive the relation 
 \begin{equation}
 f^\mathrm{b}_{N_\mathrm{H}}  \simeq 
{
\int_{0}^{\infty}
  \mathcal{B}_{m_\ell}  ( N_\mathrm{H}, Z) \,
f_{Z}  \, d\, Z
 \over 
\int_{0}^{\infty}
f_{Z}  \, d\, Z
 } ~  f_{N_\mathrm{H}} ~~.
 \label{fb_H}
\end{equation}

In a similar way, by integrating  Eq. (\ref{HZdistr}) in $d N_\mathrm{H}$, 
we obtain the   relation 
\begin{equation}
 f^\mathrm{b}_{Z}  \simeq   
{
\int_{N_\mathrm{DLA}}^{\infty} 
  \mathcal{B}_{m_\ell}  ( N_\mathrm{H}, Z) \,
f_{N_\mathrm{H}}  \, d\, N_\mathrm{H} 
 \over 
\int_{N_\mathrm{DLA}}^{\infty} 
f_{N_\mathrm{H} }  \, d\, N_\mathrm{H}
 } ~  f_{Z} ~~,
 \label{fb_zeta}
\end{equation}
where $N_\mathrm{DLA}=10^{20.3}$ atoms cm$^{-2}$.

\subsubsection{Condition of validity of Eq. (\ref{approximation}) }

We start by
  deriving the relation between the true number
of detectable DLAs in absence of obscuration,
$g_{m_\ell}(N_\mathrm{H},Z,z) \, d N_\mathrm{H} \, dZ\, dz$,
and the number observed when all sources of obscuration are considered,
$g_{m_\ell}^\mathrm{obs}(N_\mathrm{H},Z,z) \, d N_\mathrm{H} \, dZ\, dz$.

At redshift $z$ the observed number equals the true number  
minus the fraction of DLAs 
missed  as a consequence of their own extinction, 
$\varphi = \varphi_{m_\ell} (N_\mathrm{H},Z, z) $,
minus the fraction of DLAs at redshift $z$ 
missed due to the extinction of other DLAs at $z^\prime < z$
that lie in the same line of sight, $\varphi^\prime
=\varphi_{m_\ell}^\prime (N_\mathrm{H},Z, z)$.
This implies
\begin{equation}
g_{m_\ell}^\mathrm{obs}(N_\mathrm{H},Z,z)  =
g_{m_\ell}(N_\mathrm{H},Z,z) \times
\left( 1 - \varphi  -
\varphi^\prime \right) \, .
\label{A1}
\end{equation}
%
 In the term $\varphi^\prime$ 
 we should not count the DLAs
at redshift $z$ that are missed due to their own extinction, since they
are already counted in the term $\varphi$.
We should instead count only the fraction 
$\left( 1 - \varphi \right)$
of DLAs at redshift $z$ that are not missed
due to their own extinction.
Of this residual fraction, we must in turn take the fraction
of DLAs at redshift $z$ that have an additional DLA at $z^\prime < z$  
in the same line of sight, $F(z^\prime,z)$.
Of the resulting fraction, we must finally consider 
the mean fraction of DLAs at redshift $z^\prime < z$ that are
missed due to their own extinction, $\overline{\varphi}
(z^\prime \!\! < \!z)$. 
From these considerations we obtain the relation 
\begin{equation}
\varphi^\prime =
\left( 1 - \varphi \right) \times F(z^\prime,z)
\times \overline{\varphi}(z^\prime \!\!  < \!z) . 
\label{A2}
\end{equation}

From Eqs. (\ref{philost}) and (\ref{HZzdistr}) we have 
$ 
g_{m_\ell}^\mathrm{b}(N_\mathrm{H},Z,z) =
g_{m_\ell}(N_\mathrm{H},Z,z) \times
\left( 1 - \varphi 
\right)$. Comparing with  Eq. (\ref{A1})
one can see that the approximation of
 Eq. (\ref{approximation}),
$g_{m_\ell}^\mathrm{obs}(N_\mathrm{H},Z,z) 
\simeq
g_{m_\ell}^\mathrm{b}(N_\mathrm{H},Z,z) $,
 is valid when
\begin{equation}
\varphi^\prime \ll 
1- \varphi 
\label{A3}
\end{equation}

By combining Eq. (\ref{A2}) with  (\ref{A3}) we obtain that
\begin{equation}
F(z^\prime,z) \times \overline{\varphi}_{m_\ell}(z^\prime \!\! < \!z) 
 \ll 1
\label{A4}
\end{equation}
describes the limits of validity of the approximation
(\ref{approximation}). 
We estimate the validity of this condition at $z \simeq 2.3$, the mean
absorption redshift of the surveys. 

From a recent summary (Ellison et al. 2004)  of the unbiased number density of DLAs,
$n_\mathrm{DLA}(z)$, we calculate
that the probability of intersecting 2 DLAs in a random line of sight
up to $z \simeq 2.3$ is $\approx 0.15$. 
We take this value as  an estimate of $F(z^\prime,z)$.

By comparing the number densities of biased and unbiased surveys
there is room for no more than $\approx 50\%$ of missed DLAs at 
$z \simeq 2.3$ (Ellison et al. 2001, 2004). This represents an upper limit
to $\overline{\varphi}_{m_\ell}(z^\prime< 2.3) $ because the  
extinction in the observer's frame tends to decrease
with decreasing redshift and therefore
the missed fraction ${\varphi}_{m_\ell}(z^\prime<z)$ tends to diminish
at lower redshifts. 

All together, we expect therefore
$F(z^\prime,z) \times \overline{\varphi}_{m_\ell}(z^\prime<z) \la 0.08$
at $z \simeq 2.3$ and lower values  
at lower redshifts. 
We conclude that the approximation (\ref{approximation}) is good, even if it may slightly  underestimate   the
obscuration fraction.

\subsubsection{Test of Eqs.  (\ref{BigBHZz}),  (\ref{HZzdistr}) and  (\ref{fb_H}) }

\begin{figure}
   \centering 
     \includegraphics[width=7cm,angle=0]{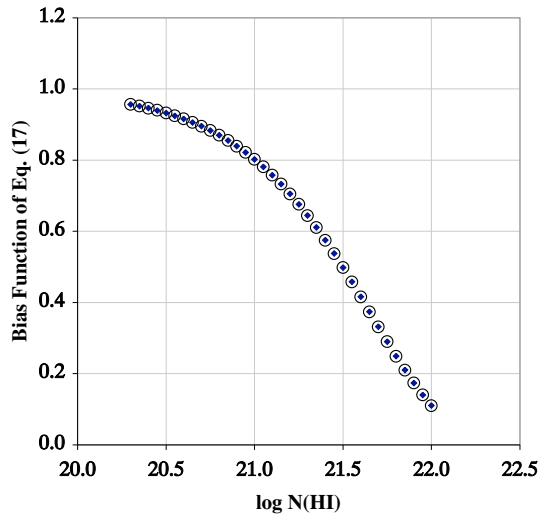}  
  \caption{ A test of validity of the equations  
  presented in Section 4.1. 
Circles: 
predictions of our Eq. (\ref{BigBHZz}).
Diamonds: 
$f_o(N,z)/f_t(N,z)$ predicted from Eq. (21) by Fall \& Pei (1993).
The adopted parameters are given in Appendix A.1.2.
  }
 \label{figAppB}%
    \end{figure}

To test  our formulation we  cross-checked its predictions in a particular case considered by Fall \& Pei (1993).  Namely, we compared the predictions of our Eq. (\ref{HZzdistr}), calculated at constant metallicity $Z$, 
with those of Eq. (21) by FP93, which is calculated at a constant 
dust-to-gas ratio $k$. This test is equivalent to insert in Eq.   (\ref{fb_H})
a Dirac delta distribution of metallicities. 
To make this comparison we transformed the power-law distribution of quasar luminosities 
used by FP93 [their Eq. (8)] into a magnitude distribution  $n(m;z)$
for our Eq. (\ref{BigBHZz}). 
We adopted $\beta=2$  and $\xi({\lambda \over 1+z})=2.78$ as common values in both equations. 
In Eq. (21; FP93) we adopted a value of  $k=0.04$ consistent with our choice of 
$G=1$, 
$Z/Z_{\sun}=0.1$ and $f_\mathrm{Fe} (Z)=0.52$ 
for this particular test. 
In Fig. \ref{figAppB} we compare the resulting ratio
$f_o(N,z)/f_t(N,z)$ predicted from Eq. (21) by FP93  
with the corresponding ratio
$g_{m_\ell}^\mathrm{b}(N_\mathrm{H},Z,z)/
g_{m_\ell}(N_\mathrm{H},Z,z)$ predicted from our Eq.  (\ref{HZzdistr}),
which equals the term $B_{m_\ell}  ( N_\mathrm{H}, Z,z)$ of
Eq.  (\ref{BigBHZz}). Our predictions
(empty circles) perfectly match those obtained with the equations
derived by FP93 (filled diamonds). 
%
Our results
are independent of $m_\ell$ provided
$m_\ell \gg 1$ and $m_\ell \gg A_\lambda$.
For  $m_\ell \approx 19$ - $20$ 
these conditions are well  satisfied. 

%

\subsection{The quasar magnitude distribution}

We derive the relation
between     the "true" magnitude distribution of quasars, $n(m; z)$, 
and the observed one, which is biased by the obscuration effect.
We consider only the obscuration due to DLA systems. 
%
%
We start by defining
$q^\mathrm{b} (m,  z_e) \, dm  \, dz_e $
the observed (biased)
number  of quasars  with   magnitude  $m \in (m,m+dm)$
at redshift $z_e \in (z_e, z_e+d z_e)$.
We then divide the quasars according to the number $i$
of DLAs that     lie in their direction.
We call $\mathcal{F}_i(z_e)$ the fraction of quasars at redshift $z_e$
that have $i$ intervening DLAs. 
Depending on the value of $i$, the distribution
of apparent magnitudes of the quasars will be different. 
We call $q^\mathrm{b}_i (m,  z_e) \, dm  \, dz_e $
the biased number of quasars  with $i$ foreground DLAs,
  magnitude  $m \in (m,m+dm)$
and redshift $z_e \in (z_e, z_e+d z_e)$.
%
%
We now assume that 
the fraction of quasars with two or more  
  DLAs   is negligible (see A.1.1), i.e. that $\mathcal{F}_{i\geq 2}(z_e) \ll \mathcal{F}_0(z_e)$
and $\mathcal{F}_{i\geq 2}(z_e) \ll \mathcal{F}_1(z_e)$. 
%
%
From the condition 
$\sum_{i=0}^{\infty} \mathcal{F}_i(z_e)=1$, we obtain
$\mathcal{F}_0(z_e) \simeq 1-\mathcal{F}_1(z_e)$.
This gives
\begin{equation}
q^\mathrm{b}(m, z_e)
\simeq 
\left[ 1-\mathcal{F}_1(z_e) \right] \, q(m,z_e)  
+ \mathcal{F}_1(z_e) \, q^\mathrm{b}_1(m, z_e)  \, 
 \label{nbprime_m}
\end{equation}
where $q(m,z_e) = q^\mathrm{b}_0(m, z_e) $  
because the true distribution, $q(m,z_e)$, must equal the
distribution of quasars without foreground DLAs,
$q^\mathrm{b}_0(m, z_e)$.
 
We now integrate Eq. (\ref{nbprime_m}) in $dz_e$ over the   interval
$(z ,z_e^\mathrm{max})$, where $z$
is the absorption redshift and  $z_e^\mathrm{max}$
the maximum redshift of the quasars in the survey. 
We assume   $\mathcal{F}_1(z_e)$ to
vary smoothly in the interval, so that it can be
approximated with its   mean value 
$\overline{\mathcal{F}}_1$. 
%
This gives
\begin{equation}
n^\mathrm{b}(m; z)
\simeq 
\left[ 1-\overline{\mathcal{F}}_1 \right] \, n(m; z) 
+ \, \overline{\mathcal{F}}_1 \, n^\mathrm{b}_1(m;z) ~,
 \label{nb_m}
\end{equation}
where  
$n^\mathrm{b}(m; z)
\equiv \int_{z}^{z_{e}^\mathrm{max}} q^\mathrm{b} (m,  z_e) \, d z_e$
and
$n^\mathrm{b}_1(m; z)
\equiv \int_{z}^{z_{e}^\mathrm{max}} q^\mathrm{b}_1 (m,  z_e) \, d z_e$.
%

We now derive an independent relation
for $n^\mathrm{b}_1(m;\overline{z})$. 
%
The apparent magnitude of the  quasars with one DLA,
 $m$, equals   the
sum of the  "true"   magnitude in absence of extinction, $m^\prime$,
plus the DLA extinction in the observer's frame,  $A_\lambda$.
The frequency distribution of the sum $m =m^\prime+A_\lambda$
is given by the convolution of the frequency distributions
of $m^\prime$ and $A_\lambda$, since these are independent variables. 
This implies
\begin{equation}
n^\mathrm{b}_1(m; z) = K 
\int_{-\infty}^{\infty} 
n(m^\prime; z) ~ 
f_{A_\lambda}\!(m-m^\prime;  \overline{z}_e) ~ dm^\prime ~,
\label{p_m}
\end{equation}
where    $K$ is a normalization factor,
$f_{A_\lambda}(A_\lambda; \overline{z}_e)$ 
is the frequency
distribution of the extinctions 
of all the DLAs in the redshift range $z \in (0,\overline{z}_e)$,
and $\overline{z}_e$
is the mean redshift of the quasars in the survey,
$\overline{z}_e \in (z ,z_e^\mathrm{max})$.

The fraction  
$\overline{\mathcal{F}}_1$ 
 is estimated
from  statistical surveys.  From the statistics 
of quasars with $\overline{z}_e  \simeq 3.0$ one finds,
in the redshift path  observable from ground ($z \gsim 1.8$,
$\overline{z} \simeq 2.3$),
   that $\approx 40\%$   of the quasars have one single DLA  
 (see P\'eroux et al. 2003).  
The  true fraction 
$\overline{\mathcal{F}}_1$ will be slightly higher than $0.4$
owing to the limitations of the wavelength coverage of the spectra. 
We adopt here $\overline{\mathcal{F}}_1 \simeq 0.5 \pm 0.1$.

We   assume that $f_{A_\lambda}(A_\lambda; \overline{z}_e)$
can be approximated with the distribution of extinctions
in the redshift range where most DLAs are observed,
$z \in (1.8,3.0)$. 
In this way the problem of estimating 
$f_{A_\lambda}(A_\lambda; \overline{z}_e)$ is brought back
to that of estimating the  distributions of column densities and metallicities
in the same redshift range, $f_{N_\mathrm{H} }$ and $f_{Z}$.
Assuming that 
$N_\mathrm{H}$ and $Z$  are statistically independent variables,
on the basis of Eq. (\ref{AlamHzeta})
we obtain the extinction distribution by convolving 
$f_{N_\mathrm{H} }$ and $f_{Z}$.

The normalization factor $K$  
is determined 
from the condition $\int   n^\mathrm{b}_1(m; z) \, dm = 
\int   n(m; z) \, dm$,
which, from the integration in $dm$ of Eq. (\ref{nb_m}), is equivalent
to 
$\int   n^\mathrm{b}(m; z) \, dm = 
\int   n(m; z) \, dm$.
These conditions guarantee that the total number of quasars
is  independent of the bias, since the only
effect of the extinction is a redistribution of the  
apparent magnitudes.

%


%

%



\begin{thebibliography}{}

\bibitem[]{} Akerman, C.J., Ellison, S. L., Pettini, M., \& Steidel, C.C.
2005, \aap, 440, 499


\bibitem[]{} Bergeson, S. D., \& Lawler, J.E.	1993, \apj, 408, 382

\bibitem[]{} Bohlin, R.C., Savage, B.D., \& Drake, J.F.
1978, \apj, 224, 132

\bibitem[]{} Blanton, M.R., Hogg, D.W., Bahcall, N.A., Brinkmann, J., Britton, M.,
et al. 2003, \apj, 592, 819

\bibitem[1998]{B&98} Boiss\'e, P., Le Brun, V., Bergeron, J., \& Deharveng, J.M.
1998, \aap, 333, 841

\bibitem[]{} Cardelli, J.A., Clayton, G.C., \& Mathis, J.S.
1988, \apj, 329, L33 (CCM)

\bibitem[]{} Cen, R., Ostriker, J.P., Prochaska, J.X., \& Wolfe, A.
2003, \apj, 598, 741

\bibitem[]{} Churches, D.K., Nelson, A.H., \& Edmunds, M.G.
2004, \mnras, 347, 1234  


\bibitem[]{} Cuesta-Bolao, M.J., \& Serna, A. 2003, \aap, 405, 917

\bibitem[]{} Dickey, J.M., \& Lockman, F.J. 1990, \araa, 28, 215



\bibitem[]{} Draine, B.T. 2003, \araa, 41, 241

\bibitem[]{} Ellison, S. L., Churchill, C.W., Rix, S.A., \& Pettini, M.
2004, \apj, 615, 118

\bibitem[]{} Ellison, S. L., Yan, L., Hook, I.M., Pettini, M., Wall, J.V., \& Shaver, P.
2001, \aap, 379, 393

\bibitem[]{} Fall, S.M. \& Pei, Y. 1989, \apj, 337, 7 

\bibitem[]{} Fall, S.M. \& Pei, Y. 1993, \apj, 402, 479 (FP93)


\bibitem[]{} Fukugita, M., Ichikawa, T., Gunn, J.E., Doi, M.,
Shimasaku, K., \& Schneider, D.P. 1996, \aj, 111, 1748

\bibitem[]{} Gordon, K.D., Clayton, G.C., Misselt, K.A., Landolt, A.U., \& Wolff, M.J.
2003, \apj, 594, 279

\bibitem[]{} Gratton, R., Carretta, E., Claudi, R., Lucatello, S., \& Barbieri, M.
2003, \aap, 404, 187


\bibitem[]{} Guarinos, J. 1991, in Evolution of Interstellar Matter and Dynamics of Galaxies,
eds. J. Palous, W.B. Burton, \& P.O. Lindblad (Cambridge Univ. Press, Cambridge, England), p. 149





\bibitem[2001]{HBP01} Hou, J.L., Boissier, S., \& Prantzos, N.
2001, \aap, 370, 23






\bibitem[]{} Khersonsky, V.K., \& Turnshek, D.A. 1996, \apj, 471, 657



\bibitem[]{} Lamareille, F., Mouhcine, M., Contini, T., Lewis, I., \&
Maddox, S. 2004, \mnras, 350, 396


\bibitem[]{} Le Brun, V., Smette, A., Surdej, J., \& Claeskens, J.-F.
2000, \aap, 363, 837



\bibitem[]{} Ledoux, C., Petitjean, P.,  M\o ller, P., Fynbo, J. \& Srianand, R.
2005, in Probing Galaxies through Quasar Absorption Lines, Proc. IAU Coll. No. 199,
eds. P.R. Williams et al., in press (astro-ph/0504402)


\bibitem[]{} Ledoux, C., Petitjean, P.,  \& Srianand, R. 2003, \mnras, 346, 209


%


\bibitem[]{} M\'enard, B., \& P\'eroux, C. 2003, \aap, 410, 33


\bibitem[]{} Mishenina, T.V., Kovtyukh, V.V., Soubiran, C., Travaglio, C., 
\& Busso, M. 2002, \aap, 396, 189



 
\bibitem[]{} Murphy, M.T., \& Liske, J. 2004, \mnras, 354, L31  

\bibitem[]{} Nagamine,ÊK., Springel,ÊV., \& Hernquist,ÊL. 2004, \mnras,
348, 435
 
\bibitem[]{} Nissen, P.E., Chen, Y.Q., Asplund, M., \& Pettini, M.
2004, \aap, 415, 993

\bibitem[]{} Ostriker, J.P., \& Heisler, J. 1984, \apj, 278, 1

\bibitem[1995]{pf95} Pei, Y.C., \& Fall, S. M. 1995, \apj, 454, 69 (PF95)

\bibitem[1991]{pfb91} Pei, Y.C., Fall, S. M.,  \& Bechtold, J. 1991, \apj, 378, 6

\bibitem[1991]{pfb91} Pei, Y.C., Fall, S. M.,  \& Houser, M.G. 
1999, \apj, 522, 604


\bibitem[]{} P\'eroux, C., McMahon, R.G., Storrie-Lombardi, L.J., \& Irwin, M.J.
2003, \mnras, 346, 1103

\bibitem[]{} Petitjean, P., Webb, J.K., Rauch, M., Carswell, R.F., \& Lanzetta, K.
1993, \mnras, 262, 499

\bibitem[]{} Pettini, M., Ellison, S.L., Steidel. C.C., Bowen, D.V.
1999, \apj, 510, 576

\bibitem[]{} Pettini, M., Smith, L.J., Hunstead, R.W., \& King, D.L.
1994, \apj, 426, 79

\bibitem[]{} Pettini, M., Smith, L.J.,   King, D.L., \& Hunstead, R.W.
1997, \apj, 486, 665

\bibitem[]{} Poli, F., Giallongo, E., Fontana, A., Menci, N., Zamorani, G.,
et al. 2003, \apj 593, L1

\bibitem[]{} Prantzos, N. \& Boissier, S. 2000, \mnras, 315, 82


\bibitem[]{} Prochaska, J.X., Gawiser, E., Wolfe, A.M., Castro, S., Djorgovski, S.G.
2003, \apj, 595, L9


%

\bibitem[]{} Prochaska, J.X., \& Wolfe, A.M. 2002, \apj, 566, 68

\bibitem[]{} Rao , S.M., \& Turnshek, D.A., 2000, \apjs, 130, 1

\bibitem[]{} Roth, K.C., \& Blades, J.C. 1995, \apj, 445, L95

\bibitem[]{} Russell, S. C., \& Dopita, M. A. 1992, \apj, 384, 508 

\bibitem[]{} Savage, B.D., Massa, D., Meade, M., \& Wesselius, P.R.
1985, \apjs, 59, 397

\bibitem[]{} Savaglio, S., Fall, S.M., \& Fiore, F. 2003, \apj, 585, 638

\bibitem[]{} Scalo, J., \& Lazarian, A. 1996, \apj, 469, 189


\bibitem[]{} Schaye, J. 2001, \apj, 562, L95

\bibitem[]{} Schechter, P. 1976, \apj, 203, 297

\bibitem[]{} 
Schneider, D.P., Fan, X., Hall, P.B., Jester, S., Richards, G.T., et al.
2003, \aj, 126, 2579



\bibitem[]{} Smette, A., Claeskens, J.F., \& Surdej, J.
1997, New Astronomy, 2, 53



\bibitem[]{} Spitzer, L. 1978, {\em Physical Processes in the Interstellar Medium},
(New Yor: Wiley Interscience)


\bibitem[]{} Storrie-Lombardi, L.J., \& Wolfe, A. M. 2000, \apj, 543, 552

\bibitem[]{} Tytler, D. 1987, \apj, 321, 49


\bibitem[]{} Vanden Berk, D.E., Quashnock, J.M., York, D.G., \& Yanny, B.
1996, \apj, 469, 78

\bibitem[]{} Van Steenberg, M.E., \& Shull, J.M.
1988, \apjs, 67, 225


\bibitem[2002]{V02b} Vladilo, G. 2002, \aap, 391, 407

\bibitem[2002]{V02b} Vladilo, G. 2004, \aap, 421, 479

%



\bibitem[2001]{VCBH01} Vladilo, G., Centuri\'on, M., Bonifacio, P., \& Howk, J.C.
2001a, \apj, 557, 1007

\bibitem[]{} Vladilo, G., Molaro, P. \& Centuri\'on, M.
2001b, in The Birth of Galaxies, Xth Rencontres de Blois (1998),
Eds. B. Guiderdoni et al., p. 425

\bibitem[]{} Wyse, R.F.G., \& Gilmore, G. 1995, \aj, 110, 2771


\bibitem[]{} Wolfe, A.M., Gawiser, E., \& Prochaska, J.X.
2003, \apj, 593, 235

\bibitem[]{} Wolfe, A.M., Lanzetta, K.M., Foltz, C.B., Chaffee, F.H.
1995, \apj, 454, 698

\bibitem[]{} Wolfe, A.M., Turnshek, D.A., Smith, H.E., \& Cohen, R.D.
1986, \apjs, 61, 249

\end{thebibliography}
\end{document}